\documentclass{aa}

\usepackage{graphicx}
\usepackage{txfonts}
\usepackage{ulem}  
\usepackage{booktabs}  

\def \ga{\mathrel{\mathchoice   {\vcenter{\offinterlineskip\halign{\hfil
$\displaystyle##$\hfil\cr>\cr\sim\cr}}}
{\vcenter{\offinterlineskip\halign{\hfil$\textstyle##$\hfil\cr
>\cr\sim\cr}}}
{\vcenter{\offinterlineskip\halign{\hfil$\scriptstyle##$\hfil\cr
>\cr\sim\cr}}}
{\vcenter{\offinterlineskip\halign{\hfil$\scriptscriptstyle##$\hfil\cr
>\cr\sim\cr}}}}}

\begin{document}

   \title{ NH$_{3}$ (1,1) hyperfine intensity anomalies in the Orion A molecular cloud}

   \author{Dong-dong Zhou \inst{1}\fnmsep\inst{3}
          \and
          Gang Wu \inst{1}\fnmsep\inst{2}
          \and
          Jarken Esimbek \inst{1}\fnmsep\inst{2}
          \and
          Christian Henkel \inst{4}\fnmsep \inst{5}\fnmsep\inst{1}
          \and
          Jian-jun Zhou \inst{1}\fnmsep\inst{2}
          \and
          Da-lei Li \inst{1}\fnmsep\inst{2}
          \and
          Wei-guang Ji \inst{1}
          \and
          Xing-wu Zheng \inst{6}
          }

   \institute{Xinjiang Astronomical Observatory, CAS  150, Science 1-Street Urumqi, Xinjiang 830011, PR China\\
              \email{wug@xao.ac.cn}
              \and
              Key Laboratory of Radio Astronomy, Chinese Academy of Sciences, Urumqi 830011, PR China
              \and
              University of Chinese Academy of Sciences, No.19(A) Yuquan Road, Shijingshan District, Beijing, 100049, P.R.China
              \and
              Max-Planck-Institut f\"{u}r Radioastronomie, Auf dem H\"{u}gel 69, 53121, Bonn, Germany
              \and
              Astronomy Department, King Abdulaziz University, PO Box 80203, Jeddah 21589, Saudi Arabia
              \and
              School of Astronomy and Space Science, Nanjing University, 163 Xianlin Avenue, Nanjing 210023, PR China
             }

   \date{Received September 15, 1996; accepted March 16, 1997}


  \abstract
{Ammonia (NH$_{3}$) inversion lines with their numerous hyperfine components are a commonly used tracer in studies of Molecular Clouds (MCs). In Local Thermodynamical Equilibrium (LTE), the two inner satellite lines (ISLs)   and the two outer satellite lines (OSLs) of the NH$_{3}$($J,K$) = (1,1) transition are each predicted to have equal intensities.
However, hyperfine intensity anomalies (HIAs) are observed to be omnipresent in star formation regions, which is still not fully understood.
In addressing this issue, we find that the computation method of the HIA by the ratio of the peak intensities may have defects, especially when being used to process the spectra with  low velocity dispersions. Therefore we define the integrated HIAs of the ISLs (HIA$_{\rm IS}$) and OSLs (HIA$_{\rm OS}$) by the ratio of their redshifted to blueshifted  integrated intensities (unity implies no anomaly) and develop a  procedure to calculate them. Based on this procedure, we present a systematic study of the integrated HIAs in the northern part of the Orion A MC. We find that integrated HIA$_{\rm IS}$ and HIA$_{\rm OS}$ are commonly present in the Orion A MC and no clear distinction is found at different locations of the MC. The medians of the integrated HIA$_{\rm IS}$ and HIA$_{\rm OS}$  are 0.921$\pm$0.003 and 1.422$\pm$0.009, respectively, which is consistent with the HIA core model and inconsistent with the collapse or expansion (CE) model. Selecting those 170 positions where both integrated HIAs deviate by more than 3-$\sigma$ from unity, most (166) are characterized by HIA$_{\rm IS}$ < 1 and HIA$_{\rm OS}$ > 1, which suggests that the HIA core model plays a more significant role than the CE model. The remaining four positions are consistent with the CE model. We compare the integrated HIAs with the para-NH$_{3}$ column density ($N$(para-NH$_{3}$)), kinetic  temperature ($T_{\rm K}$), total velocity dispersion ($\sigma_{\rm v}$),  non-thermal velocity dispersion ($\sigma_{\rm NT}$), and the total opacity of the NH$_{3}$($J,K$) = (1,1) line ($\tau_{0}$).
The integrated HIA$_{\rm IS}$ and HIA$_{\rm OS}$ are almost independent of $N$(para-NH$_{3}$).
The integrated HIA$_{\rm IS}$ decreases slightly from unity (no anomaly) to about 0.7 with increasing $T_{\rm K}$, $\sigma_{\rm v}$, and $\sigma_{\rm NT}$ .
The integrated HIA$_{\rm OS}$ is independent of $T_{\rm K}$ and reaches values close to unity with increasing $\sigma_{\rm v}$ and $\sigma_{\rm NT}$ .
The integrated HIA$_{\rm IS}$ is almost independent of  $\tau_{0}$, while the integrated HIA$_{\rm OS}$ rises with $\tau_{0}$ showing then higher anomalies.
These correlations can not be fully explained by neither the HIA core nor the CE model.}

 \maketitle

%

\section{Introduction}
\label{intro}

Ammonia (NH$_{3}$) is a very important tracer in determining the density \citep[e.g.][]{1969ApJ...157L..13C}, temperature \citep[e.g.][]{1981ApJ...246..761H}, velocity \citep[e.g.][]{1993ApJ...406..528G}, and intrinsic line width \citep[e.g.][]{1998ApJ...504..207B} of Molecular Clouds (MCs).
Due to its high abundance \citep[for the observed range, see][]{1983ApJ...270..589B, 1987A&A...173..352M, 2001ApJ...554L.143W},
specific hyperfine structure and sensitivity to kinetic temperature \citep{1983ARA&A..21..239H}, NH$_{3}$ inversion transitions are commonly used in studies related to star forming regions (SFRs), MCs, and nearby galaxies \citep[e.g.][]{1998ApJ...505L.151Z, 2005A&A...440..893H, 2014ApJ...790...84L,2018A&A...616A.111W}.
The ground state para NH$_{3}$ ($J,K$) = (1,1) transition consists (due to electric quadrupole  splitting) of five distinct components, namely the main component ($\Delta F$=0) and four satellite components ($\Delta F$=$\pm$1), two on each side of the main component \citep{1983ARA&A..21..239H}.
Weaker magnetic spin-spin interactions introduce a total of  18 hyperfine components within the profiles of the five quadrupole hyperfine components \citep[see Fig.1 in ][]{1977ApJ...215L..35R}.

The two inner satellite lines (ISLs) and the two outer satellite lines (OSLs) are predicted to have equal intensities (26\% for each ISL and 22\% for each OSL with respect to the intensity of the main line under conditions of LTE and optically thin emission). This is used to model important parameters such as opacity, temperature, and column density \citep[e.g.][]{1983ARA&A..21..239H}.
However, the expectation of equal intensities of the ISLs and OSLs was found to be not always valid. \citet{1977ApJ...214L..67M} discovered a hyperfine intensity anomaly (HIA) in the case of absorption spectra towards the continuum background of DR21.
Later, HIAs were found to be commonly present in MCs and SFRs \citep[e.g.][]{1984A&A...139..258S, 2007MNRAS.379..535L, 2012PASJ...64...30N}.
Recently \citet{2015ApJ...806...74C} studied the HIA in a sample of 343 SFRs. They found the HIA is ubiquitous in high-mass SFRs, and also found there is no clear correlation between the HIA and the temperature, line width,  optical depth, or the stage of stellar evolution.
However, the result may be affected by the computation method these authors have used to calculate the HIA (see Section \ref{results}).  Therefore, the aim of this article is to quantify such uncertainties and to propose a better way to determine HIAs.

What is causing HIAs? In analytical and numerical calculations, the HIA can be reproduced by non-LTE populations induced by trapping of selected hyperfine transitions
in the rotational lines of NH$_{3}$ connecting the ($J,K$) = (2,1) and (1,1) inversion doublets \citep[e.g.][]{1977ApJ...214L..67M}.
An alternative scenario involves systematic motions like expansion and contraction \citep[e.g.][]{2001A&A...376..348P}.

If the former mechanism causes the HIA, the emitting
cloud  should be composed of several small  cores with line widths of about 0.3 -- 0.6 km s$^{-1}$ (hereafter core model), since clouds with larger line widths would reshuffle the non-thermal populations \citep[see][]{1985A&A...144...13S}. The HIA is then expected to decrease with  density, temperature, and line width \citep[e.g.][]{2015ApJ...806...74C}.
The HIA of the ISLs (redshifted versus blueshifted component,  HIA$_{\rm IS}$) can only be smaller than unity while the HIA of the OSLs (again redshifted versus blueshifted component, HIA$_{\rm OS}$ ) can only be larger than unity \citep[e.g.][]{ 1985A&A...144...13S}.
Based on this mechanism, the HIA emitting clouds should be composed of many small (10$^{-2}$ pc), high density (10$^{6}$ -- 10$^{7}$ cm$^{-3}$) clumps of 0.3 -- 1 M$_{\odot}$ \citep[e.g.][]{1977ApJ...214L..67M, 1985A&A...144...13S}, which
is further interpreted as being compatible with the "competitive accretion" model of high-mass star formation \citep[e.g.][]{2015ApJ...806...74C}.

In the latter case, (systematic) collapse or expansion, e.g. outflow, infall, and/or rotation in clouds, could lead to HIAs (hereafter CE model). Photons emitted from one hyperfine transition can be absorbed by another one due to systematic motions resulting in severe changes in the level populations of the NH$_{3}$ (1,1) sub-levels.
Expansion (contraction) can only strengthen the emission on the red (blue) side, while suppressing those on the other side \citep[e.g.][]{2001A&A...376..348P}. In principle a cloud with collapsing and also expanding parts may lead to emission enhancements on different sides of the ISLs and  OSLs, but such a situation may not be common \citep[e.g.][]{1977ApJ...214L..67M}.
Based on this, the HIA is expected to be a tracer of systematic motions \citep[e.g.][]{2001A&A...376..348P, 2007MNRAS.379..535L}.
It is also expected to be strengthened as the ammonia column density increases, since photon trapping processes are more efficient at larger optical depths
\citep[e.g.][]{2001A&A...376..348P}.
However, until now the HIA is still not fully understood. We still do not know why HIAs are so common in MCs and SFRs. There is even no clear correlation between the HIA and the physical properties of a cloud \citep{1984A&A...139..258S, 2015ApJ...806...74C}.

A systematic observational study of  HIAs could help us to address their origin. However, in previous observational studies, the authors focused on single point observations towards separate star formation regions. Based on the latest extended and sensitive Green Bank Ammonia Survey\footnote{https://greenbankobservatory.org/science/gbt-surveys/gas-survey} (GAS) \citep{2017ApJ...843...63F}, we firstly provide a systematical study of the HIA in an extended and prominent  cloud, the Orion A MC.
With the criteria outlined in Section \ref{results}, we selected 1383 spectra in the northern part of the Orion A MC and study the morphology and statistics of the HIA as well as correlations between the HIA and several molecular gas parameters derived from these observations.


\section{Data}
\label{data}

The archival NH$_{3}$ data we used to calculate the HIAs are derived from Green Bank Telescope (GBT) observations of the northern part of the Orion A MC which is a part of the Green Bank Ammonia Survey \citep{2017ApJ...843...63F}.
The seven-beam K-Band Focal Plane Array (KFPA) was used as the frontend and the VErsatile GBT Astronomical Spectrometer (VEGAS) was used as the  backend.
A spectral resolution of 5.7 kHz (0.07 km s$^{-1}$ at 23.7 GHz) is derived under the VEGAS Mode 20.
At the observed frequency, the GBT has a primary beam (FWHM) of about 32" (0.065 pc). The data were observed in OTF mode and resampled with a sample step of about 10".  For more details, see \citet{2017ApJ...843...63F}.

The Orion A MC is the nearest and probably best studied MC that continues to produce both low- and high-mass stars \citep{2008hsf1.book..459B}. At a distance of about 414 pc \citep{2007A&A...474..515M}, it can be observed with
good linear resolution. Meanwhile it is located about 15 degrees below the Galactic
plane, leading to a less confused background than that typically
encountered along the Galactic plane. The Orion A MC is the largest
MC in the Orion complex \citep{2005A&A...430..523W}. The GBT observed region covers the northern compact ridge of the Orion A MC, which is an integral-shaped filamentary cloud \citep{1987ApJ...312L..45B, 1999ApJ...510L..49J, 2006ApJ...653..383J, 2017ApJ...843...63F, 2018A&A...616A.111W}.

Baseline removal is  critical when analysing NH$_{3}$ data, because any error in the slope of a baseline may lead to fake HIAs.
In order to minimize this problem, we only used linear polynomials to remove the baselines and all of these spectra selected have flat and gently varying baselines which can be well fitted by the linear polynomials.
The large threshold of the signal-to-noise ratio (SNR) we used (see below) will likely further reduce any baseline related effects.

Matplotlib\footnote{https://matplotlib.org} \citep{2007MCSE..9..90}, lmft\footnote{https://lmfit.github.io/lmfit-py} \citep{2014LMFIT..2014},  scipy\footnote{https://www.scipy.org}, ApLpy\footnote{https://aplpy.readthedocs.io} \citep{2001SciPy...2001}, montage\footnote{http://montage.ipac.caltech.edu}  and GILDAS\footnote{http://www.iram.fr/IRAMFR/GILDAS/}  software packages were used in the data reduction and display.

\begin{figure*}[!htbp]
\centering
\includegraphics[width=0.48\textwidth]{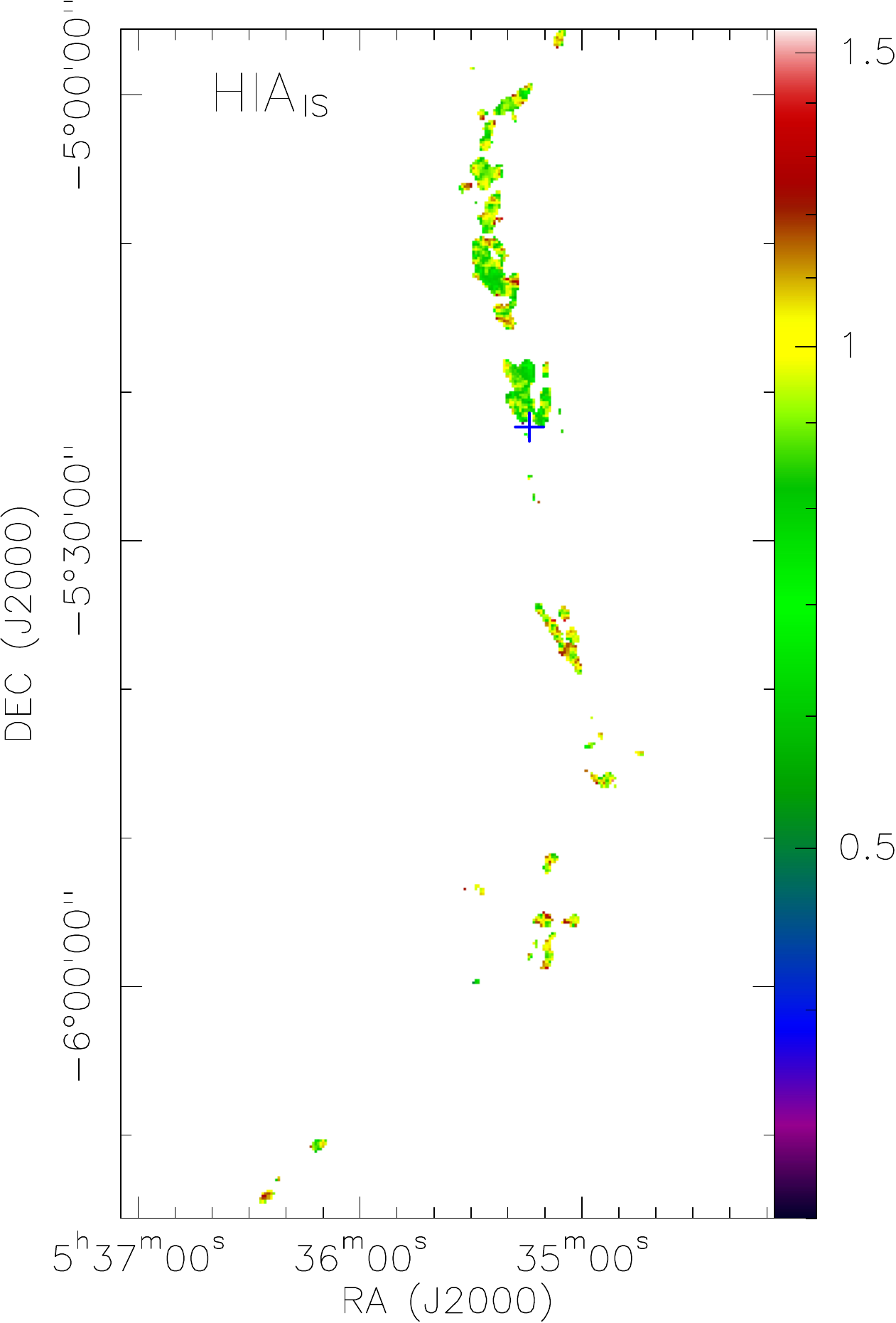}
\includegraphics[width=0.48\textwidth]{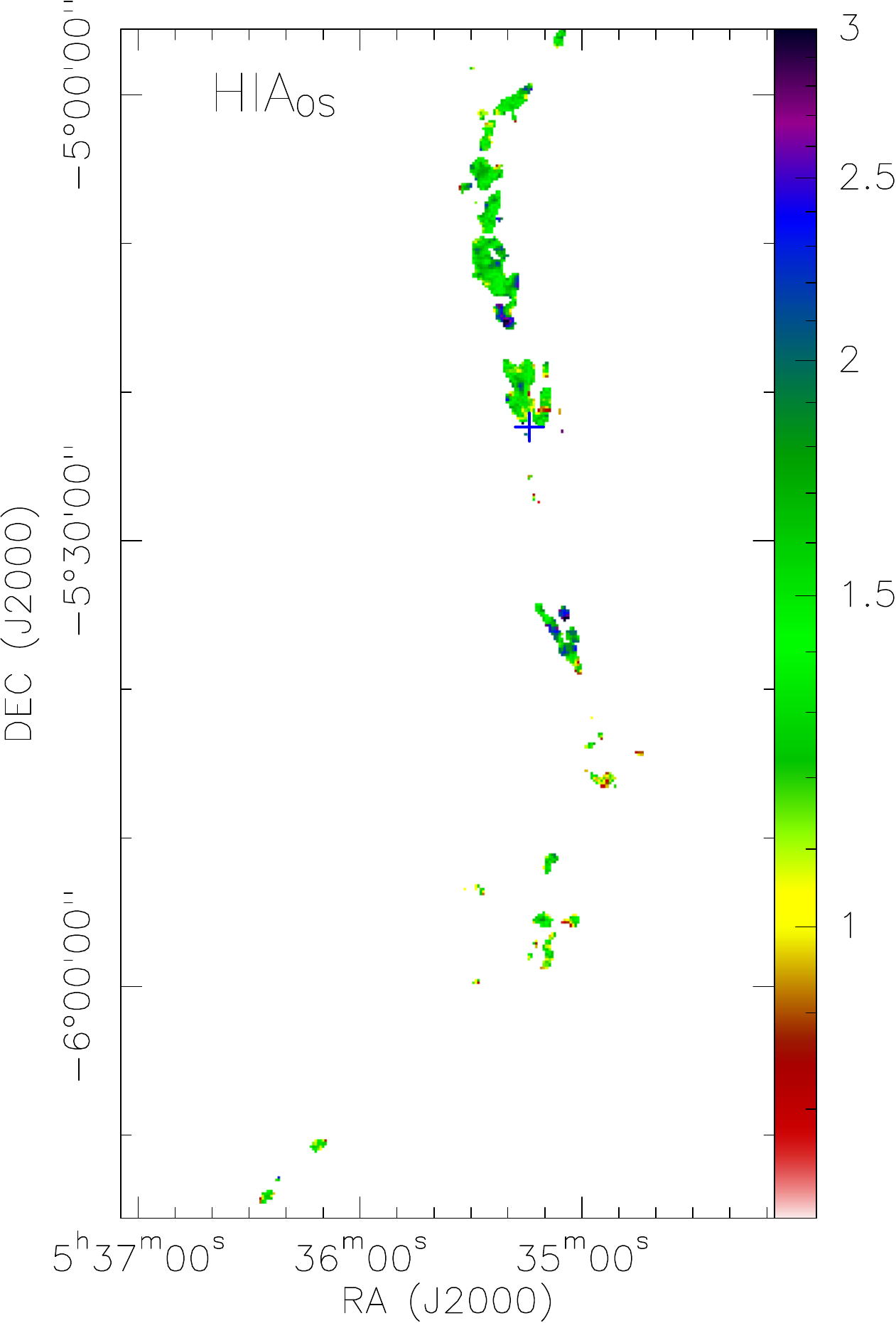}
\caption{The integrated hyperfine intensity anomalies of the inner satellite lines (left panel) and the outer satellite lines (right panel) derived from 1383 spectra with an signal-to-noise ratio larger than 15 in the Integral-Shaped Filament (ISF) of the Orion A molecular cloud. The blue cross in each panel illustrates the location of the Orion Kleinmann-Low Nebula (R.A. 05:35:14.16 DEC -05:22:21.5 J2000). }
\label{fig:hia_dis}
\end{figure*}

\section{Results}
\label{results}

As we have mentioned in the introduction, there is a total of 18 hyperfine components within the profiles of the five quadrupole hyperfine components of the ($J,K$) = (1,1) line.
The ISLs (OSLs) contain three (two) hyperfine components with different velocity separations \citep[see the cyan vertical lines in Fig. \ref{fig:mspec} and also ][]{1977ApJ...215L..35R}.
Assuming all the 18 hyperfine components are optically thin and have Gaussian profiles
with the same velocity dispersion, as we may expect, the ISLs (or OSLs) which are the combination of Gaussian spectra of three (or two) hyperfine components with different offsets should have different line widths and peak intensities.
To the contrary, the  integrated intensities (flux) of the ISLs (or OSLs) over specific integrated velocity ranges should show equal values (see Fig. \ref{fig:simulation} in the Appendix).
Therefore, {\it the HIA calculated with the peak intensities from single Gaussian fittings}  \citep[hereafter peak HIA; e.g.][]{2007MNRAS.379..535L, 2015ApJ...806...74C} {\it does not reflect the real anomaly. The HIA calculated from integrated intensities} (hereafter integrated HIA) {\it provides a less biased view} (see Appendix \ref{computation} for details).

Therefore, to calculate the HIA in this paper, we adopt

 \begin{equation}
    HIA_{\rm IS} =  \frac{F_{\rm RISL}}{F_{\rm BISL}}\\
  \end{equation}
and
 \begin{equation}
    HIA_{\rm OS} = \frac{F_{\rm ROSL}}{F_{\rm BOSL}},\\
  \end{equation}
\noindent
where $F_{\rm RISL}$ and $F_{\rm BISL}$ are the integrated intensities of the red-  and blueshifted sides of the ISLs, respectively. Accordingly $F_{\rm ROSL}$ and $F_{\rm BOSL}$ are the integrated intensities of the red-  and blueshifted sides  of the OSLs. The integrated ranges are all set to
$\pm 2 \times \Delta V$ of each of the 18 hyperfine components, which is explained in Appendix \ref{hiaPro}.

\subsection{Distributions of integrated HIAs in the Orion MC}
\label{distribution}

Based on our procedure (Appendix \ref{hiaPro}), we firstly excluded spectra with an SNR less than 15. We further excluded the spectra with  velocity dispersions larger than 1.0 km s$^{-1}$ (35 spectra mostly located around the  Kleinmann-Low (KL) Nebula in OMC 1, R.A. 05:35:14.16 DEC -05:22:21.5 J2000). In cases where the velocity dispersion is larger than 1.0 km s$^{-1}$, the main line and ISLs start to seriously overlap. Spectra with more than one velocity component were also discarded.
Finally 1383 spectra are selected to calculate the HIAs.
The total optical depths of NH$_{3}$(1,1) ($\tau_{0}$, see Appendix \ref{hiaPro}) of the selected spectra range from 0.12 to 2.98 and 89\% of them are smaller than unity.
This means for all of the selected spectra that the individual hyperfine components of the four satellite lines and even the blended emission of each inner and outer satellite group of hyperfine components is optically thin (see Table \ref{tab:nh3ratio}, where relative intensities are given, normalized to unity).
Figure \ref{fig:hia_dis} presents the distributions of the integrated HIAs calculated from the GBT data with our procedure. The left panel displays the integrated HIA$_{\rm IS}$ and the right panel displays the integrated HIA$_{\rm OS}$.
With the threshold of the SNR larger than 15,
the majority of the available pixels are distributed north of the Kleinman-Low nebula and only a small number of the pixels are scattered along the southern part of the filamentary MC.

Firstly we can see in the two panels (for the statistical results, see Section \ref{statistic}), that integrated HIA$_{\rm IS}$ and  HIA$_{\rm OS}$ are commonly present in the Orion A MC. Secondly, there is no clear trend of the integrated HIAs along the MC. For example, there is no clear distinction between the integrated HIAs of the cloud around the KL Nebula and Trapezium cluster (i.e. the OMC1; Orion Cloud 1) \citep{1996Natur.382..139W}, the more quiescent northern part of the cloud (OMC 2,3) \citep{1999ApJ...510L..49J, 2013ApJ...768L...5L}, and the more diffuse southern part of the cloud (OMC 4, 5) \citep{2006ApJ...653..383J}. Finally, the integrated HIA$_{\rm OS}$ is distributed in a wider range than the integrated HIA$_{\rm IS}$.

\subsection{Statistics of the integrated HIAs}
\label{statistic}

\begin{figure}
\includegraphics[width=\hsize]{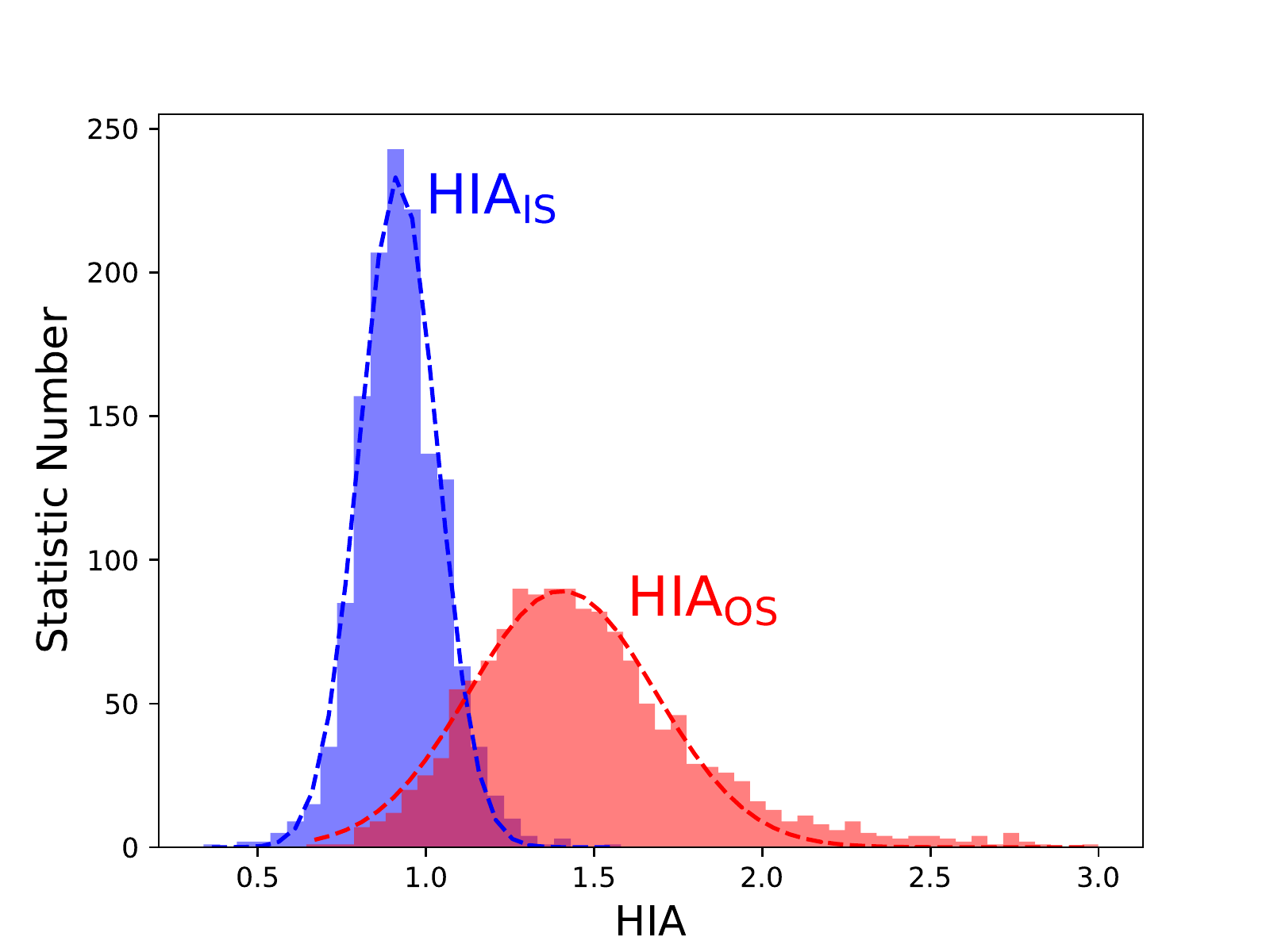}
\caption{Histograms of the integrated hyperfine intensity anomalies of the inner satellite lines (the blue histogram) and the outer satellite lines (the red histogram) of the spectra with a signal-to-noise ratio larger than 15. The dashed blue and red lines represent the fitted results by assuming that they are subject to Gaussian distributions.}
\label{fig:stat}
\end{figure}

We present the statistical results of the integrated HIAs  in Fig. \ref{fig:stat} and Table \ref{tab:hiaStat}.
We can see that, as already outlined in Section \ref{distribution},
(1) most of the ratios do not equal unity and that (2) the integrated HIA$_{\rm IS}$ is distributed in a narrower range than the HIA$_{\rm OS}$. The medians of the integrated HIA$_{\rm IS}$ and HIA$_{\rm OS}$ are 0.921$\pm$0.003 and 1.422$\pm$0.009 respectively,  the errors representing standard deviations of the mean.
The statistical results can be compared with those of  \citet{2015ApJ...806...74C} in a sample of 334 high-mass SFRs (0.889$\pm$0.004 and 1.232$\pm$0.006), which will be further discussed in Appendix \ref{intANDpeak}. (3) The integrated HIA$_{\rm IS}$ and HIA$_{\rm OS}$  are distributed on both sides of unity.

As outlined  in the introduction, the HIA core model predicts that the HIA$_{\rm IS}$ can only be smaller than unity and the HIA$_{\rm OS}$ can only be larger than unity \citep[e.g.][]{1985A&A...144...13S, 2015ApJ...806...74C}. To the contrary, the CE model predicts the HIA$_{\rm IS}$ and  HIA$_{\rm OS}$ should be simultaneously larger and smaller than unity \citep[e.g.][]{2001A&A...376..348P}.
We can see in Fig. \ref{fig:stat} that the overall 'inverse' distributions relative to unity (0.921$\pm$0.003 and  1.422$\pm$0.009, respectively) of the integrated  HIA$_{\rm IS}$ and  HIA$_{\rm OS}$ are consistent with the HIA core model \citep[e.g.][]{1977ApJ...214L..67M, 1985A&A...144...13S} and  inconsistent with the CE model  \citep[e.g.][]{2001A&A...376..348P}, as also suggested by \citet[][]{2015ApJ...806...74C}.
However, as we can see in Fig. \ref{fig:stat}, there are still some ratios of the ISLs which are larger than unity and some ratios of the OSLs which are smaller than unity.

\begin{figure}
\includegraphics[width=\hsize]{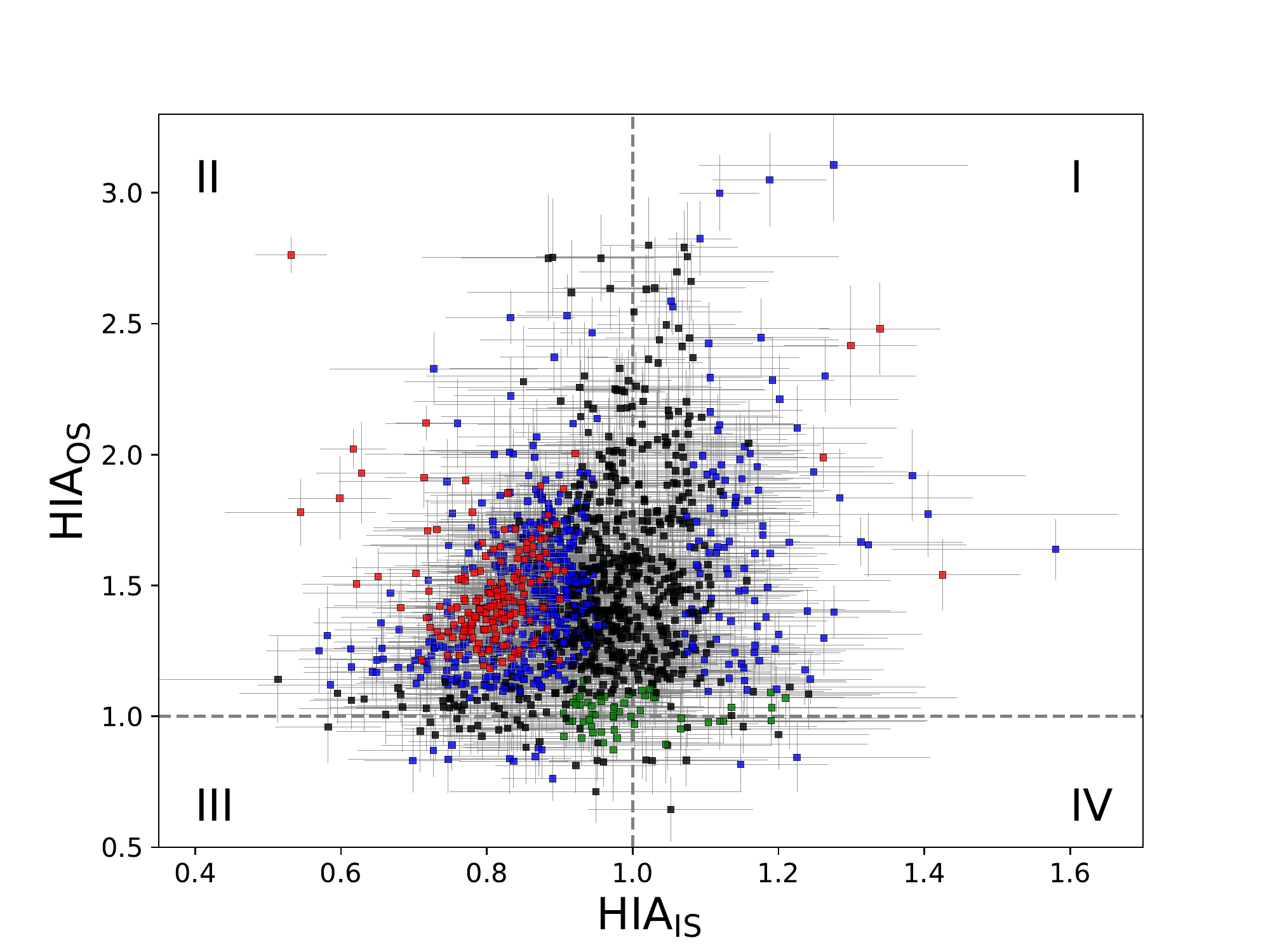}
\caption{Distribution of the integrated hyperfine intensity anomalies of the inner (HIA$_{\rm IS}$) and outer (HIA$_{\rm OS}$) satellite lines.
The red (blue) points illustrate positions, where both the HIA$_{\rm IS}$ and HIA$_{\rm OS}$ values deviate by more than 3-$\sigma$ (1-$\sigma$) from unity. Black points denote positions, where either the HIA$_{\rm IS}$ or the HIA$_{\rm OS}$ is located within 1-$\sigma$ of unity, while the green points show positions, where both the HIA$_{\rm IS}$ and HIA$_{\rm OS}$ are within 1-$\sigma$ of unity. Grey dashed vertical and horizontal lines subdivide the panel into regions with HIA$_{\rm IS}$>1 (labelled I for HIA$_{\rm OS}$>1 and IV for HIA$_{\rm OS}$<1) and HIA$_{\rm IS}$<1 (labeled II for HIA$_{\rm OS}$>1 and III for HIA$_{\rm OS}$<1).}
\label{fig:isos}
\end{figure}

To further investigate the distributions of the integrated HIAs, we present a direct plot of the  integrated HIA$_{\rm IS}$ and HIA$_{\rm OS}$ in Fig. \ref{fig:isos}.
The panel is divided into four quarters, which are labelled as I, II, III, and IV in Fig. \ref{fig:isos} by the two dashed grey lines (HIA$_{\rm IS}$=1, HIA$_{\rm OS}$=1).
The blue- and red points denote positions, where both HIA$_{\rm IS}$ and HIA$_{\rm OS}$ deviate by more than 1- and 3-$\sigma$ from unity. The black points highlight positions where only one of the HIAs deviates by more than 1-$\sigma$ from unity and the green points indicate locations, where both the HIA$_{\rm IS}$ and HIA$_{\rm OS}$ are within 1-$\sigma$ from
unity.
Again, we can see that most of the integrated HIAs do not equal unity.

The core model predicts that all points should be in the second quarter (HIA$_{\rm IS}$<1, HIA$_{\rm OS}$>1), while the CE model predicts that the points should either lie in the first quarter (HIA$_{\rm IS}$>1, HIA$_{\rm OS}$>1) or the third quarter (HIA$_{\rm IS}$<1, HIA$_{\rm OS}$<1).
We can see in the figure that most of the points are distributed in the upper two quarters (HIA$_{\rm OS}$>1), especially the second quarter, but there are also a handful of points located in the third and forth quarters.
As suggested by  e.g. \citet[][]{1984A&A...139..258S}, a deviation from unity by more than  3-$\sigma$  clearly shows anomalous emission. Taking the HIA uncertainties (the uncertainties are illustrated in Appendix \ref{hiaPro}) in consideration, there are in total 170 points, indicated by a red color,  whose integrated HIA$_{\rm IS}$ and HIA$_{\rm OS}$ deviate from unity by more than their  3-$\sigma$ uncertainties. We can see in Fig. \ref{fig:isos} that 166 red points are located in the second quarter and 4 red points in the first quarter.
These statistical properties suggest that the HIA core model is playing the dominant role, while only 4 out of 170 points support the CE model. In this context it is interesting that we find all these four positions in the first quarter of the diagram, and none in the third quarter. This is consistent, following the CE model (Sect. \ref{intro}), with expansion of the molecular gas.

\begin{table*}
\caption{The statistical results of the integrated hyperfine intensity anomaly of the inner satellite lines  and outer satellite lines.}
\label{tab:hiaStat}
\centering
\begin{tabular}{cccc}     
\hline\hline
HIA type & Median & Mean & $\sigma$ \tablefootmark{a}  \\
\hline
  The integrated HIA$_{\rm IS}$ & 0.922$\pm$0.003 & 0.926$\pm$0.003 & 0.115$\pm$0.003  \\
  The integrated HIA$_{\rm OS}$ & 1.422$\pm$0.009 & 1.475$\pm$0.009 & 0.277$\pm$0.007 \\
\hline
\multicolumn{4}{l}{\tablefoottext{a}{the deviation  by assuming a Gaussian distribution.} }
\end{tabular}

\end{table*}

\section{Discussion}

\begin{figure*}
\flushleft
\includegraphics[width=0.49\textwidth]{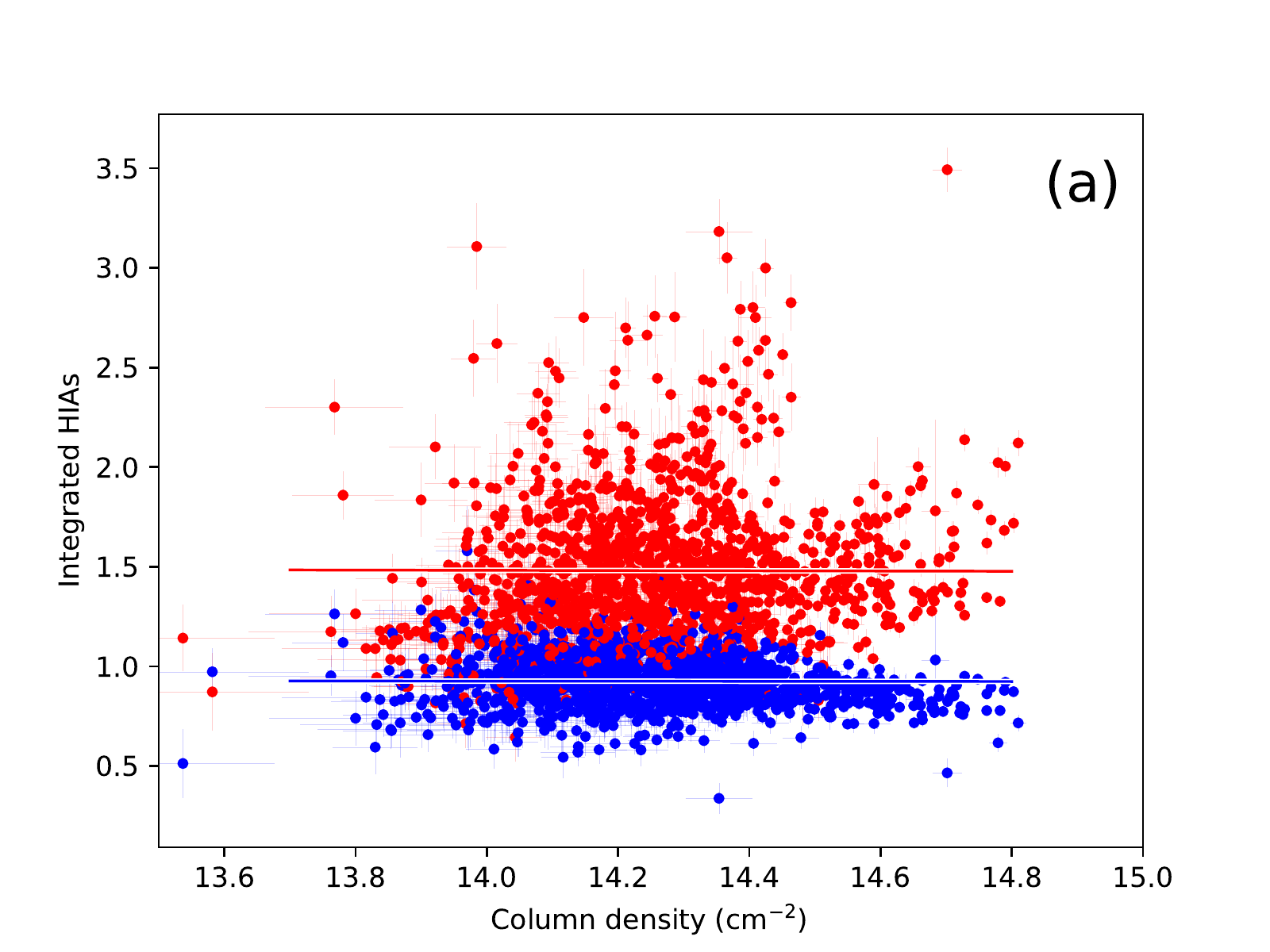}
\includegraphics[width=0.49\textwidth]{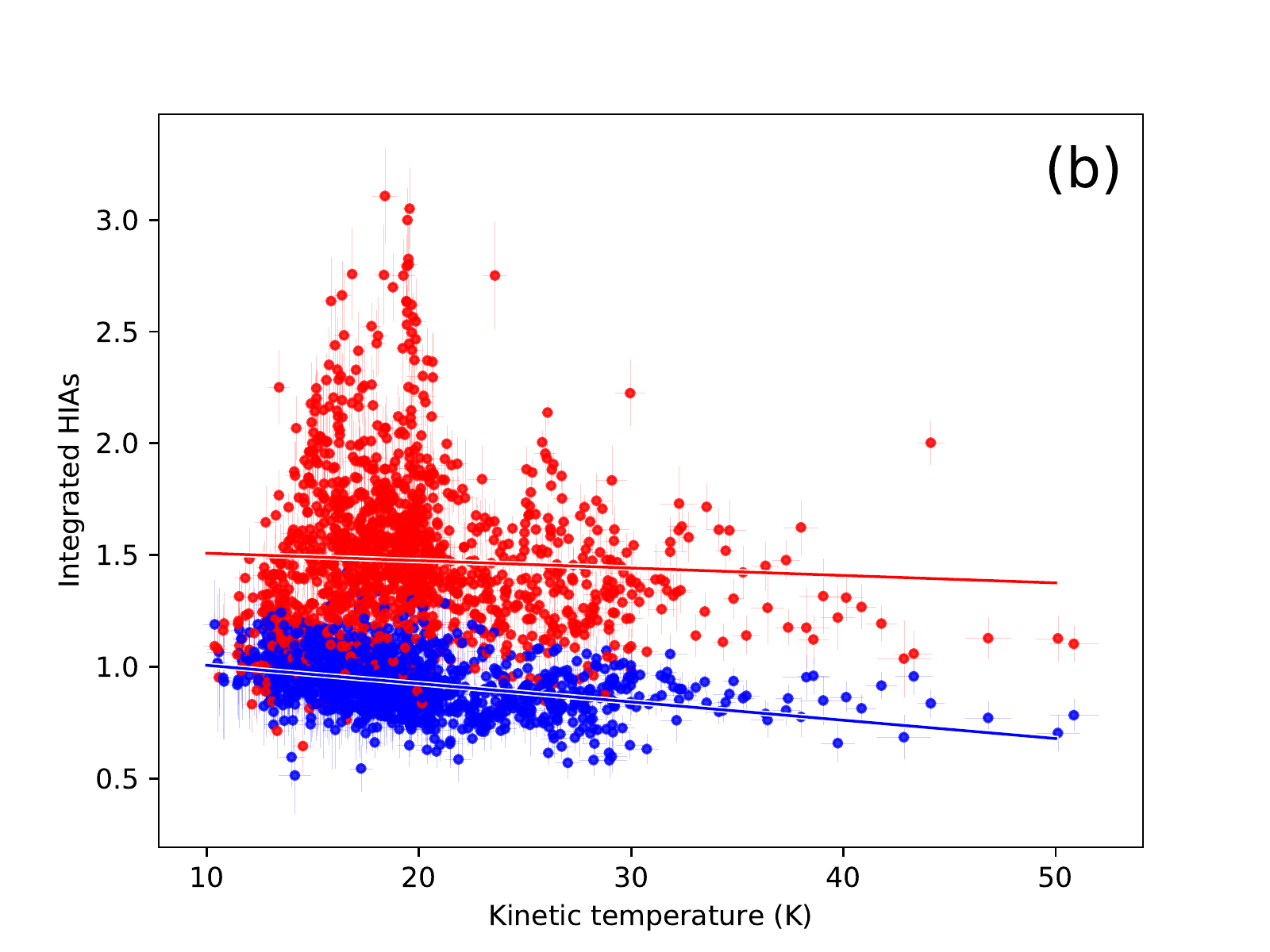}
\includegraphics[width=0.49\textwidth]{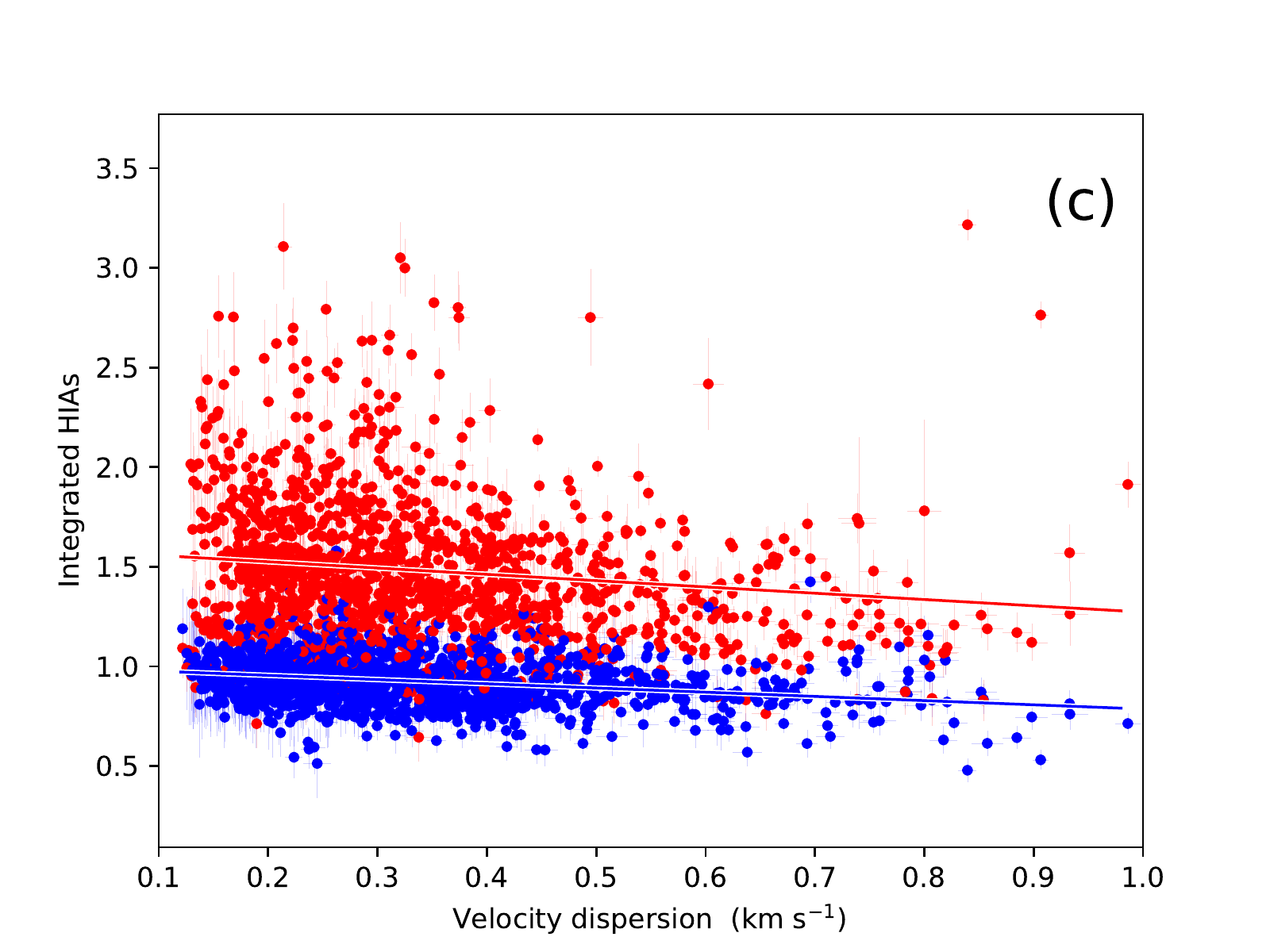}
\includegraphics[width=0.49\textwidth]{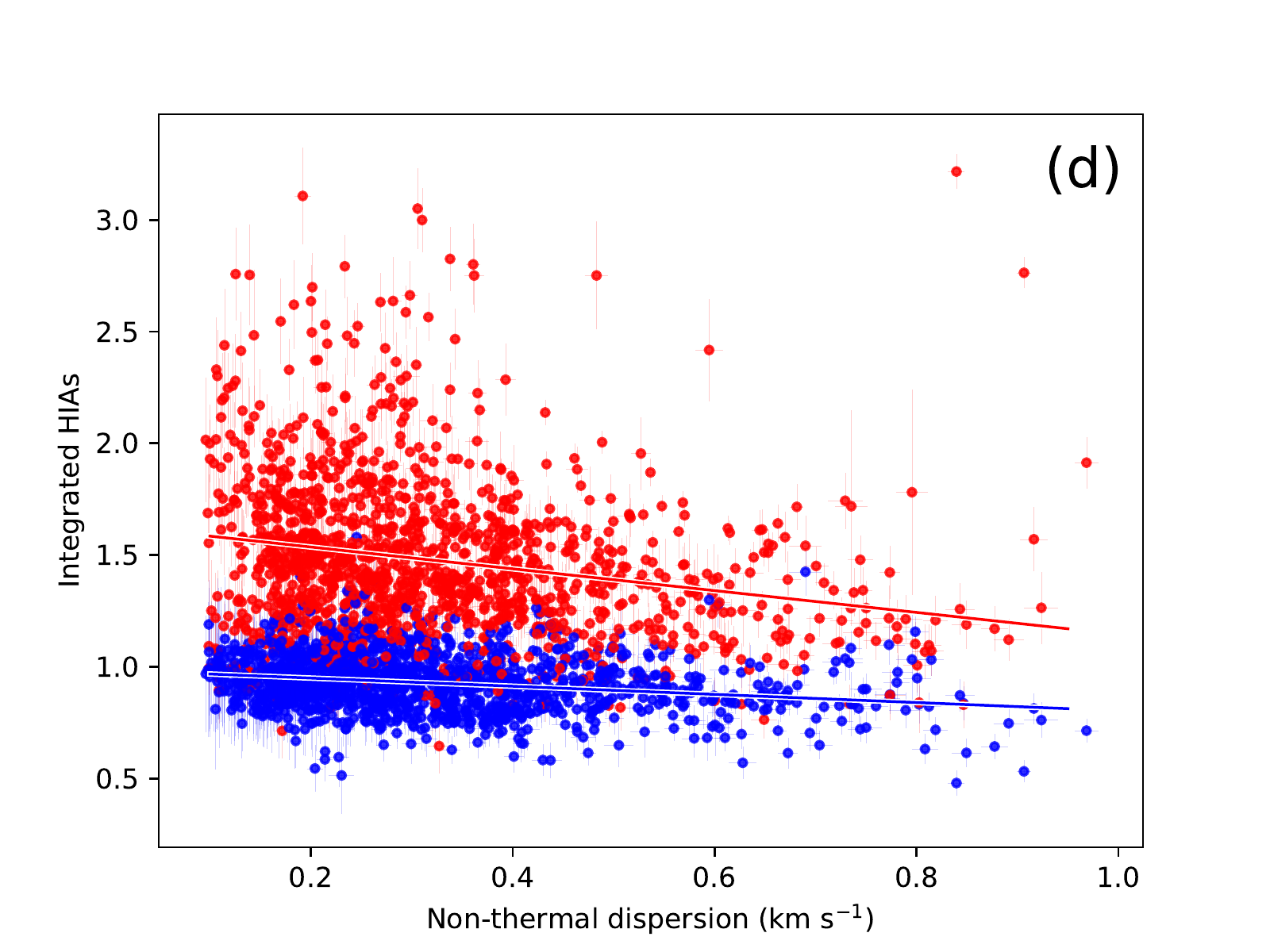}
\includegraphics[width=0.49\textwidth]{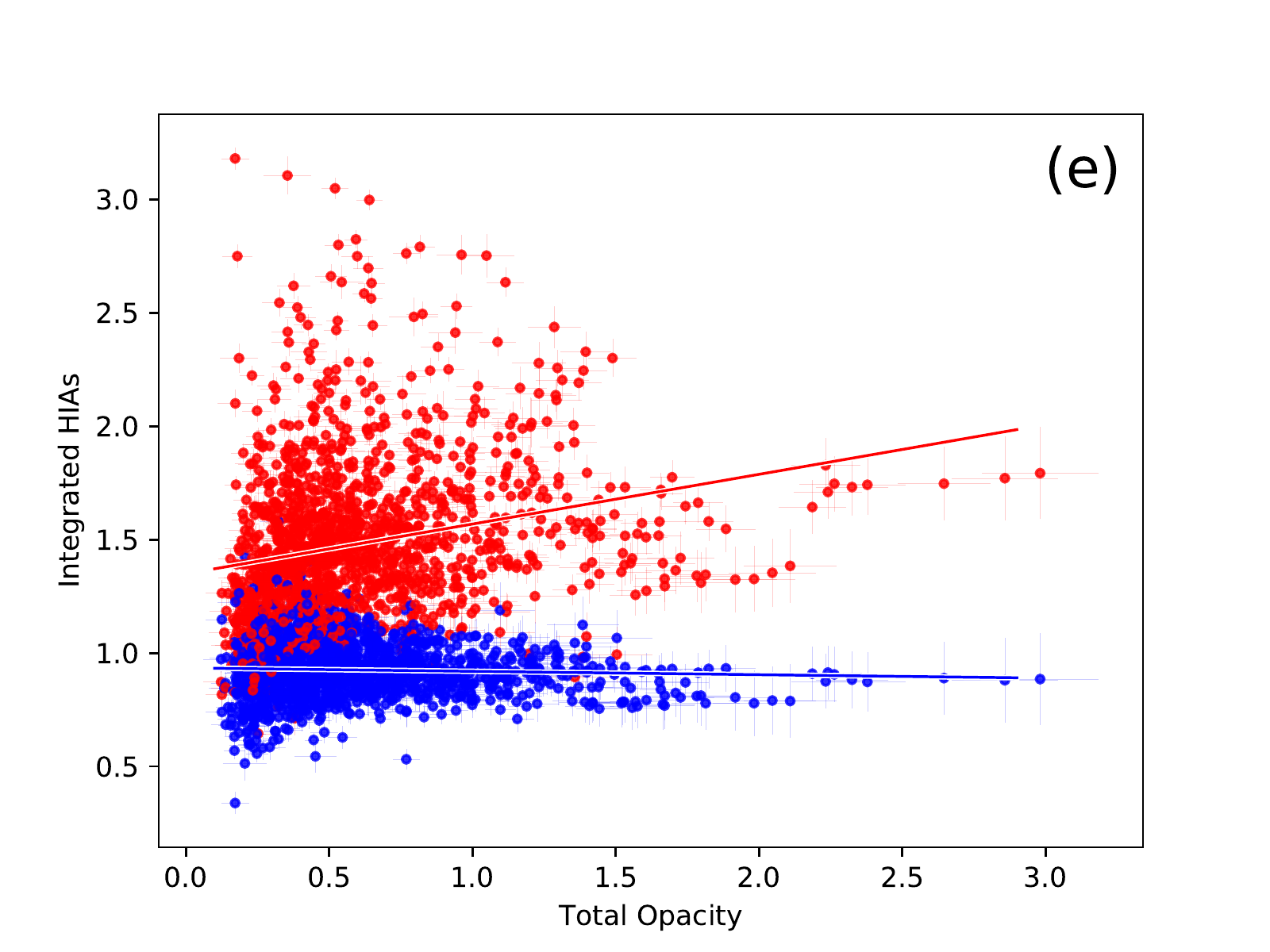}

\caption{Comparisons of the integrated hyperfine intensity anomalies (inner satellites: blue dots; outer satellites: red dots) with the para-NH$_{3}$ column density (a), kinetic  temperature  (b),  velocity dispersion (c), non-thermal velocity dispersion (d), and total optical depth of NH$_{3}$ (1,1) (e) all derived from the GBT NH$_{3}$ observations  \citep{2017ApJ...843...63F}.}
\label{fig:hia-params}
\end{figure*}

We compare the integrated HIAs with the para-NH$_{3}$ column density ($N$(para-NH$_{3}$)), kinetic  temperature ($T_{\rm K}$), intrinsic velocity dispersion ($\sigma_{\rm v}$), non-thermal dispersion ($\sigma_{\rm NT}$), and the total opacity of NH$_{3}$) (1,1) ($\tau_{0}$), which are all derived from NH$_{3}$ (1,1) and (2,2) observations. $ \sigma_{\rm v}$ and $\tau_{0}$ are derived from our procedure (Appendix \ref{hiaPro}). $N$(para-NH$_{3}$) and $T_{\rm K}$  are taken from \citet{2017ApJ...843...63F}. The routines to calculate these parameters are also illustrated there.
$\sigma_{\rm NT}$ is calculated by  $\sigma_{\rm NT} = \sqrt{\sigma_{\rm v}^{2}-k \it{T}_{\rm K} / (\it\mu_{\rm NH_{3}}  m_{\rm H} ) }$, where $\mu\rm_{NH_{3}}$ = 17,
$m\rm_{H}$ = 1.674 $\times$ 10$^{-24}$ g, k is the Boltzmann constant, and $T\rm_{K}$ is the kinetic  temperature calculated from NH$_{3}$ (1,1) and (2,2).
We summarize all the comparisons in Fig. \ref{fig:hia-params}. The linear regression results are also presented in each panel (red and blue lines) and in
 Table \ref{tab:hia-params}. In all the regressions,
the correlation coefficient takes a range of values from -1 (a perfect negative  correlation) to +1 (a perfect positive correlation). A correlation coefficient of zero indicates no relationship between the two variables being compared.

Firstly we can see from panel (a) in Fig. \ref{fig:hia-params} that the integrated HIAs are almost independent of $N$(para-NH$_{3}$) (the correlation coefficients of the integrated HIA$_{\rm IS}$ and HIA$_{\rm OS}$ are -0.029 and 0.027 respectively).
Then we can see from  panel (b) in Fig. \ref{fig:hia-params} that the integrated HIA$_{\rm IS}$ decreases from unity to about 0.7, indicating higher anomalies, with increasing temperature. However, the correlation is weak, with a correlation coefficient of -0.35.
On the other hand, the integrated HIA$_{\rm OS}$ seems to show a decrease at the high temperature end ($T_{\rm K}$ > 40 K). Probably due to the large dispersion of the integrated HIA$_{\rm OS}$, a clear correlation between the integrated HIA$_{\rm OS}$ and $T_{\rm K}$ is not present (the correlation coefficient is -0.05).
We can also see from panels (c) and (d) in Fig. \ref{fig:hia-params} that the integrated HIAs
decrease slightly with increasing $\sigma_{\rm v}$ and $\sigma_{\rm NT}$, but again with low correlation coefficients (see Table \ref{tab:hia-params}).
The integrated HIA$_{\rm IS}$ deviates more and more from unity, while the HIA$_{\rm OS}$ moves closer and closer to unity with increasing $\sigma_{\rm v}$ or $\sigma_{\rm NT}$.
It should be noted that $T_{\rm K}$, $\sigma_{\rm NT}$, and $\sigma_{\rm v}$ are not independent from each other. Thus, the correlations between the integrated HIAs , $T_{\rm K}$, $\sigma_{\rm v}$, and  $\sigma_{\rm NT}$ show similarities.
Finally, we can see from panel (e) in Fig. 4 that the integrated HIA$_{\rm IS}$ appears to be independent of $\tau_{0}$. On the contrary, the integrated HIA$_{\rm OS}$ appears to rise with  $\tau_{0}$ and shows higher anomalies with increasing $\tau_{0}$.

We should note that the correlation coefficients are all small. The maximal correlation coefficient of -0.35 is found between the integrated HIA$_{\rm IS}$ and T$_{\rm K}$.
Overall, the correlations between the integrated HIAs and potentially related parameters (Fig. 4) are weaker than what the models predict.
This may partially be due to the very large dispersion in the HIA$_{\rm OS}$.
Nevertheless, the current results are not fully explained, neither by the the HIA core model nor by the CE model.
For example, the HIA core model can not explain that the integrated HIA$_{\rm IS}$ deviates more and more from unity with increasing $T_{\rm K}$, $\sigma_{\rm v}$, and $\sigma_{\rm NT}$.
The HIA core model requires subcores with line widths of 0.3 -- 0.6 km s$^{-1}$ (Section \ref{intro}). The strength of the anomaly is then expected  to decrease with increasing temperature and line width.
The HIA CE model can not explain the integrated HIA$_{\rm IS}$ and  HIA$_{\rm OS}$  have opposite trends relative to unity with increasing  $\sigma_{\rm v}$ and $\sigma_{\rm NT}$. The integrated HIAs are independent of $N$(para-NH$_{3}$), while the CE core model predicts that the integrated HIA$_{\rm IS}$ and  HIA$_{\rm OS}$ are  strengthened when the ammonia column density increases \citep[e.g.][]{2001A&A...376..348P}.
Finally, the correlations of the integrated HIAs with $\sigma_{\rm NT}$  suggest that non-thermal motions (e.g. turbulence) might also have contributions to the HIAs as suggested by \citet{2015ApJ...806...74C}.

\begin{table*}
\renewcommand \tabcolsep{15pt}
\caption{The statistical results of the integrated hyperfine intensity anomalies (HIAs) of the inner satellite lines (ISLs) and outer satellite lines (OSLs) as function of para-ammonia column density, kinetic temperature, velocity dispersion, the turbulent part of the velocity dispersion, and the total opacity of  NH$_{3}$ (1,1).}
\label{tab:hia-params}
\centering
\begin{tabular}{lccccccc}     
\hline\hline
  parameters &  \multicolumn{3}{c}{Inner satellite lines (ISLs)}  &  &  \multicolumn{3}{c}{Outer satellite lines (OSLs)}  \\
\cline{2-4}
\cline{6-8}

  & slope\tablefootmark{a} & intercept\tablefootmark{b}  &  r\tablefootmark{c}  & & slope\tablefootmark{a} & intercept\tablefootmark{b} & r\tablefootmark{c}  \\
\hline

$N$(para-NH$_{3}$)  (cm$^{-2}$)       & -0.002             & 0.958            & -0.029           &  &  0.006             & 1.564         &  -0.027   \\
$T_{\rm K}$ (K)                       & -0.008             & 1.088            & -0.350           &  & -0.003             & 1.541         & -0.050   \\
$\sigma\rm_{V}$ (km s$^{-1}$)         & -0.211             & 0.997            & -0.263           &  & -0.316             & 1.588         & -0.140   \\
$\sigma\rm_{NT}$ (km s$^{-1}$)        & -0.182             & 0.985            & -0.224           &  & -0.488             & 1.633         & -0.215   \\
$\tau_{0}$                            &-0.015              &0.934             & -0.043           &  & -0.218             & 1.353         &  0.225   \\
%
\hline
\end{tabular}
\raggedright
\tablefoot{All the regressions are derived by the linregress procedure of SciPy.}
\tablefoottext{a}{slope of the regression line.}
\tablefoottext{b}{intercept of the regression line.}
\tablefoottext{c}{correlation coefficient.}
\end{table*}

\section{Conclusions}

\begin{enumerate}
\item Through the simulations (Fig. \ref{fig:simulation}) and GBT observations (Fig. \ref{fig:comp-m-m}), we have demonstrated that the integrated HIAs calculated from the integrated intensities should be less biased than the peak HIAs calculated from the peak intensities,
    due to the different separations of the 18 hyperfine components within the inner and outer satellite lines (ISLs and OSLs) and their asymmetric profiles.

 \item We have developed (Appendix \ref{hiaPro}) a procedure to fit the NH$_{3}$ (1,1) profiles and to calculate the integrated  and  peak HIAs. Based on this procedure, we firstly present a study of the HIAs in the Orion A molecular cloud (MC). We do not find clear differences or trends in the integrated HIA$_{\rm IS}$ and HIA$_{\rm OS}$ along the filamentary MC.

 \item The medians of the statistical results of the integrated HIA$_{\rm IS}$ and HIA$_{\rm OS}$ (defined by the integrated intensity ratios of their redshifted to blueshifted  components) have "inverse"  values relative to unity, i.e. 0.921$\pm$0.003 and 1.42$\pm$0.009, respectively, which is consistent with the HIA core model and inconsistent with the CE model.
     Accounting for the 3-$\sigma$ uncertainties of the individual positions, most (166) of the integrated HIAs are characterised by HIA$_{\rm IS}$<1 and HIA$_{\rm OS}$>1, which is again consistent with the HIA core model. However, there are also four positions with HIA$_{\rm IS}$>1 and HIA$_{\rm OS}$>1, which supports the CE model, suggesting expanding gas. These results indicate that the HIA core model plays a more significant role than the CE model.

 \item We compared the integrated HIAs with the para-NH$_{3}$ column density ($N$(para-NH$_{3}$)), kinetic  temperature ($T_{\rm K}$), velocity dispersion ($\sigma_{\rm v}$), non-thermal velocity dispersion ($\sigma_{\rm NT}$) , and NH$_{3}$ total opacity ($\tau_{0}$).
     The integrated HIAs are within the uncertainties independent of $N$(para-NH$_{3}$).
     The integrated HIA$_{\rm IS}$ deviates more and more from unity with increasing temperature while the integrated HIA$_{\rm OS}$ is almost independent of $T_{\rm K}$.
     The integrated HIAs all decrease with increasing $\sigma_{\rm v}$ and $\sigma_{\rm NT}$.
     With increasing  $\sigma_{\rm v}$ and $\sigma_{\rm NT}$, the integrated HIA$_{\rm IS}$ is getting below unity and thus more anomalous, while the integrated HIA$_{\rm OS}$ decreases to unity (no anomaly).
     The integrated HIA$_{\rm IS}$ appears to be independent of $\tau_{0}$, but the integrated HIA$_{\rm OS}$ appears to rise with $\tau_{0}$ and shows higher anomalies with increasing $\tau_{0}$.
     Neither the HIA core model nor the CE model can explain all these results.

 \item
  Overall, we find that the HIA core model, related to trapping in some hyperfine level transitions, plays a more significant role. Nevertheless, the correlation results are not fully explained, neither by the HIA core model nor by the CE model, based on the assumption of  radial motions inside the studied clouds.

\end{enumerate}

\begin{acknowledgements}
We would like to thank the anonymous referee for the useful suggestions that improved this study.
This work was funded by the National Nature Science foundation of China under grant 11433008 and the CAS "Light of West China" Program under grant Nos. 2018-XBQNXZ-B-024, and partly supported by the National Natural Science foundation of China under grant 11603063, 11973076, 11703074, and 11703073. G. W. acknowledges the support from Youth Innovation Promotion Association CAS.
This research has made use of NASA's Astrophysics Data System Bibliographic Services. This research made use of Montage. It is funded by the National Science Foundation under Grant Number ACI-1440620, and was previously funded by the National Aeronautics and Space Administration's Earth Science Technology Office, Computation Technologies Project, under Cooperative Agreement Number NCC5-626 between NASA and the California Institute of Technology.
\end{acknowledgements}


\begin{appendix}

\section{The computation method: peak  or  integrated intensity ratios?}
\label{computation}
\subsection{Comparison with the simulated spectra}
\label{method}

In order to test the expectation that the integrated HIA is less biased than the peak HIA (see Section \ref{results}) and to simulate the results we could derive from the observed data, we created a simulated spectrum with a specific velocity dispersion by assuming all the 18 hyperfine components have Gaussian profiles and the same velocity dispersion \citep[e.g.][]{2008ApJS..175..509R, 2015PASP..127..266M}. The simulated spectrum can then be described by

\begin{equation}
S(V) = \sum^{18}_{i=1} R_{\rm hfs}[i] \times A_{0} \times e^{- (V - V_{0} - V_{\rm hfs}[i])^{2} /(2*\sigma_{\rm V}^{2})},
\end{equation}
\noindent
where $S(V)$ is the simulated spectrum combining all the 18 hyperfine components, $R_{\rm hfs}[i]$ and $V_{\rm hfs}[i]$ are  the relative intensities  of individual features under the conditions of LTE and optically thin emission (see Table \ref{tab:nh3ratio}) and the relative velocities, which are given in \citet[][]{1977ApJ...215L..35R} (see Table \ref{tab:nh3ratio}). $A_{0}$ and $V_{0}$ represent the arbitrary amplitude and velocity offset  of the combined spectrum, which are set to 0.5 and zero, respectively. $\sigma_{\rm V}$ denotes the velocity dispersion for each of the 18 hyperfine components.

\begin{figure}
\centering
\includegraphics[width=\hsize]{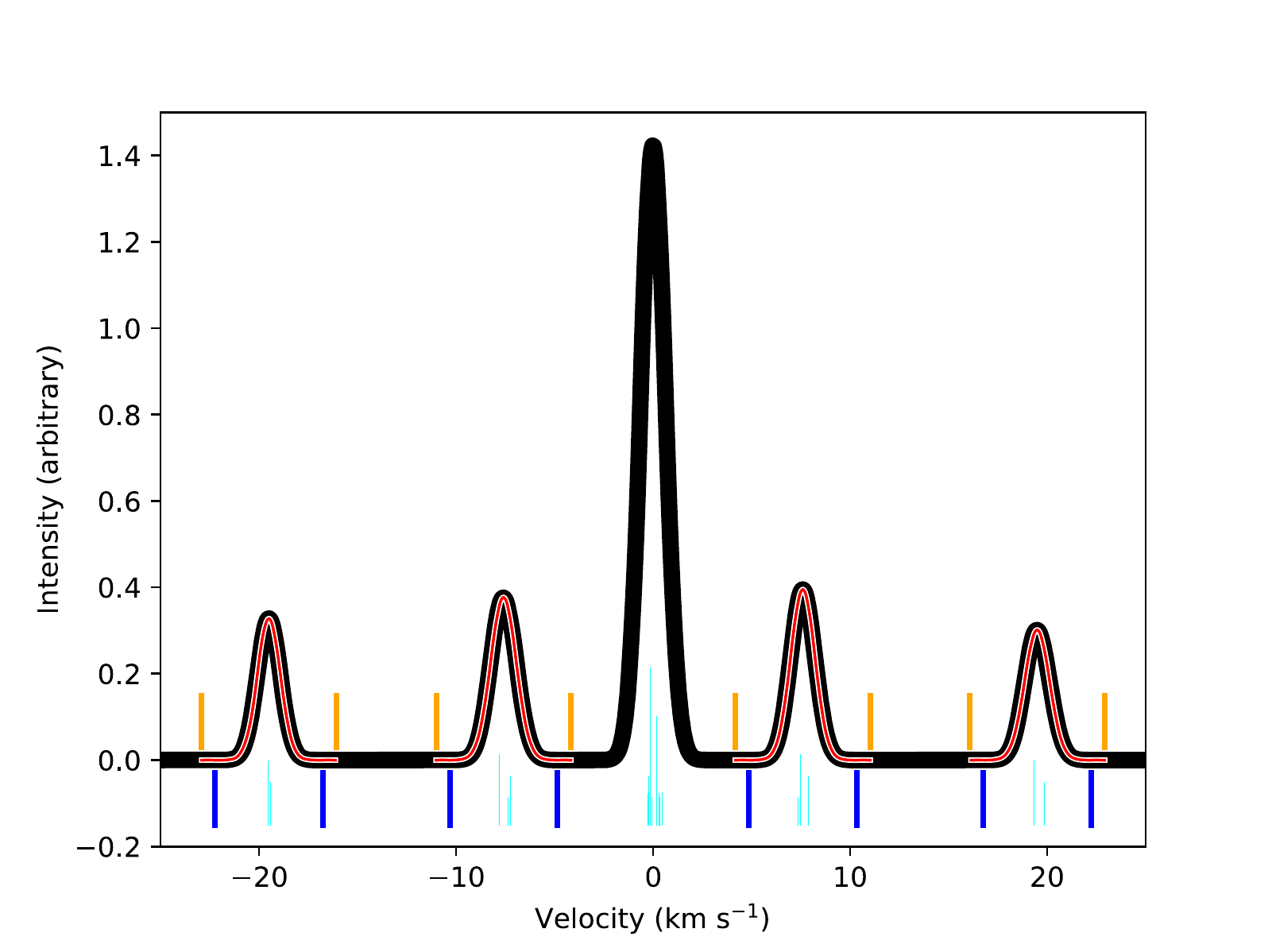}
\caption{A simulated optically thin NH$_{3}$ (1,1) spectrum under conditions of local thermodynamical equilibrium with a velocity dispersion of 0.5 km s$^{-1}$. The black line presents the simulated spectrum.
The thick orange vertical lines above the zero level show the ranges of the sub-spectra used to fit the inner and outer satellite lines with single Gaussian functions.
The thick blue vertical lines below the zero level indicate the integrated ranges used to calculate the integrated hyperfine intensity anomalies of the inner and outer satellite lines.
The red lines show the single Gaussian fitting results.  The cyan vertical lines below the zero level present the 18  individual hyperfine components listed in Table \ref{tab:nh3ratio}. The lengths and separations of the lines denote their predicted intensities and velocity separations. }
\label{fig:mspec}
\end{figure}

We present an example of a simulated spectrum with a velocity dispersion of 0.5 km s$^{-1}$ in Fig.  \ref{fig:mspec}.  The black thick line shows the combined NH$_{3}$ (1,1) spectrum.
The thick blue vertical lines below the zero level show the integrated ranges used to calculate the integrated  HIA$_{\rm IS}$ and HIA$_{\rm OS}$.
The orange vertical lines above the zero level indicate the ranges of the sub-spectra used to fit
the inner and outer satellite lines with single Gaussian functions.
The red lines present the single Gaussian fitting results.
 The cyan vertical lines present the 18 hyperfine components given in \citet{1977ApJ...215L..35R}. The lengths and the separations of these lines (see Table \ref{tab:nh3ratio}) represent their expected intensities and velocity separations.

\begin{figure}
\includegraphics[width=\hsize]{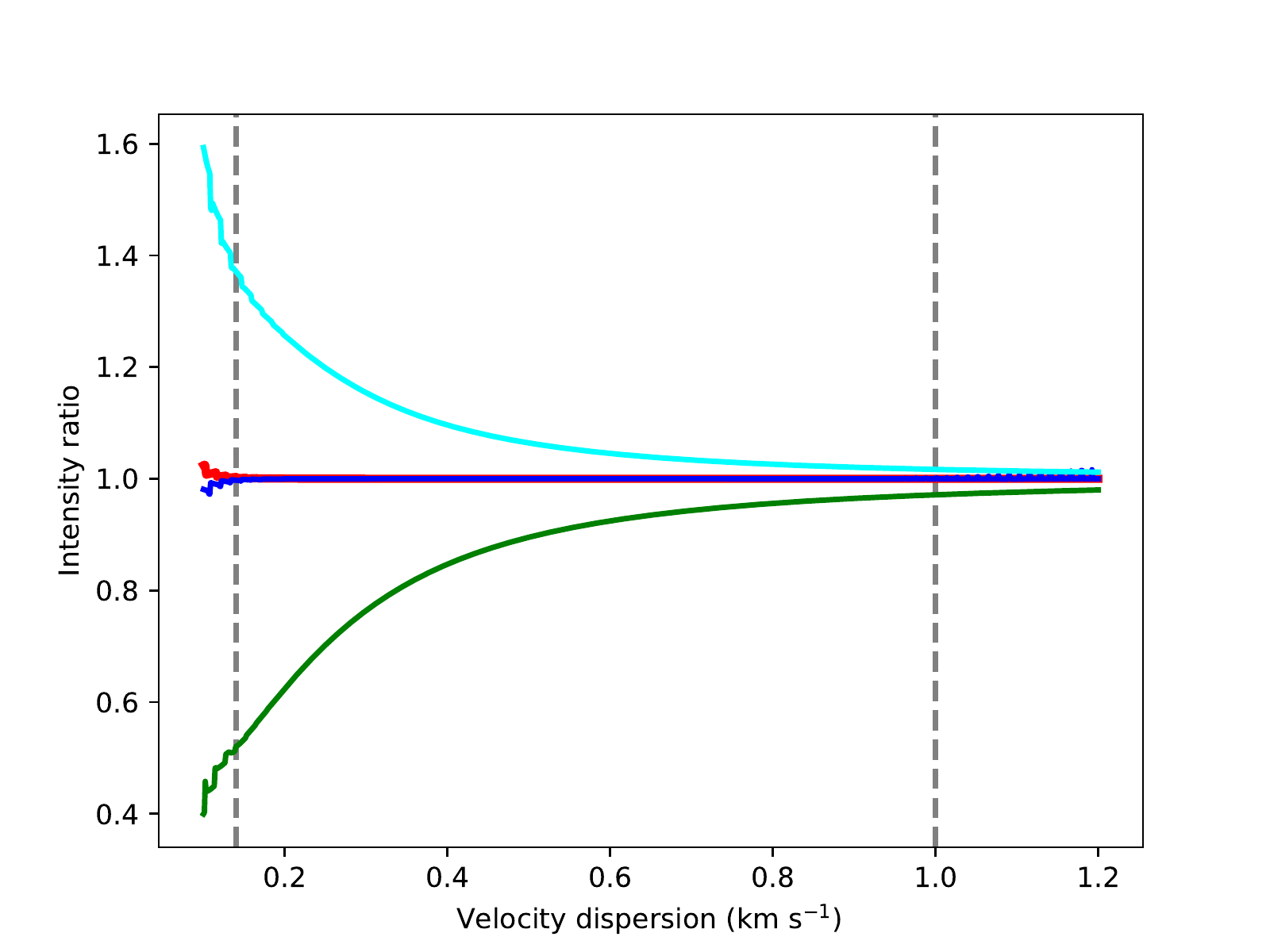}
\caption{The peak and integrated hyperfine intensity anomalies derived from the simulated spectra against the velocity dispersion. The blue and red lines present the integrated hyperfine intensity anomalies of the inner and outer satellite lines, respectively. The green and cyan lines present the peak hyperfine intensity anomalies of the inner and outer satellite lines, respectively. Two grey dashed lines indicate the velocity dispersion range of the GBT data we used. }
\label{fig:simulation}
\end{figure}

We then created 1000 simulated spectra with linearly spaced velocity dispersions from 0.1 to 1.2 km s$^{-1}$ to further study the relationship between the peak and integrated HIAs  against the velocity dispersion.
Applying the procedure which is described in Appendix \ref{hiaPro}, we fitted these spectra and derived the peak and integrated HIAs, respectively, which are presented in Fig. \ref{fig:simulation}.
The red and blue lines show the derived integrated HIA$_{\rm IS}$ and HIA$_{\rm OS}$. The green and cyan lines present the derived peak HIA$_{\rm IS}$ and  HIA$_{\rm OS}$. Two grey dashed lines indicate the velocity dispersion range of the GBT data we used  (0.14 to 1.0 km s$^{-1}$).
Oscillations of the blue line (the integrated HIA$_{\rm IS}$) are present at the large  velocity dispersion end ($\ga$1.0 km s$^{-1}$), because the main line and ISLs start to overlap and the main line has an asymmetric profile.
At the low  velocity dispersion end ($<$0.1km\,s$^{-1}$) there are fluctuations of all the HIAs.
Fluctuations of the peak HIAs arise at low velocity dispersions, because the spectra develop multi-peaks, which reduce the Gaussian fitting precision.
Fluctuations of the integrated HIAs occur because the velocity ranges at these dispersions only cover a relatively small part of the ISLs or OSLs.
However, within the velocity dispersion range of the observed data we used (dashed vertical lines in Fig. \ref{fig:simulation}), these effects can be ignored.

We should note that all the 18 hyperfine components have infinite line wings.
Obviously, this also holds for the combined five distinct quadrupole hyperfine components of the NH$_{3}$ (1,1) line.
The derived integrated HIAs  are approximately one, but are not equal to unity since the integrated range can not cover all the line wings.
For example, for velocity dispersions from 0.14 to 1.0 km s$^{-1}$ (the velocity dispersion range of the observed data we used), the maximal  deviations from unity are 0.00398 and 0.00323 for the integrated HIA$_{\rm IS}$ and HIA$_{\rm OS}$, respectively. Compared with the spectral noise of the observed data, these deviations can be neglected.

As we can see in  Fig. \ref{fig:simulation}, the integrated HIA$_{\rm IS}$ and HIA$_{\rm OS}$ are very close to unity. However, clear dichotomies are present between the peak HIAs and unity.
Therefore, with the theoretically  predicted intensities and the velocity separations of all the 18 hyperfine components, the previously used method, using the peak intensities to calculate the HIAs, can 'produce' anomalies especially at the low velocity dispersion end.
Peak and integrated HIAs derived from the observed data are compared in Appendix \ref{intANDpeak}.

\begin{sidewaystable*}[b!]
\renewcommand
\tabcolsep{4pt}
\caption{The predicted relative intensities, frequency and velocity separations of the 18 hyperfine components of NH$_{3}$ (1,1).}
\label{tab:nh3ratio}
\raggedright
\begin{tabular}{lcccccccccccccccccc}     
\hline\hline
hyperfine component number  & 1 & 2 & 3 & 4 & 5 & 6 & 7 & 8 & 9 & 10 & 11 & 12 & 13 & 14 & 15 & 16 & 17 & 18\\
\hline
   relative intensities\tablefootmark{a}  &  0.073 & 0.037 & 0.083 & 0.010 & 0.047 & 0.017 & 0.047 & 0.233 & 0.010 & 0.15 & 0.017 & 0.010 & 0.018 & 0.010 & 0.083 & 0.047 & 0.073 & 0.037\\

   frequency separations (kHz)\tablefootmark{b}  & 1545.1  & 1534.1 & 617.7 & 582.8 & 571.8 & 19.9 & 16.8 & 10.5 & 5.9 & -15.1 & -24.5 & -25.5 & -36.5 & -581.0 & -590.3 & 623.3 & -1526.8 & -1569.0 \\
   velocity separations (km s$^{-1}$)    & -19.55 &-19.41 & -7.82 & -7.37  &  -7.23  &  -0.25  &  -0.21  &  -0.13  &  -0.07  &   0.19  &   0.31  &   0.32  &   0.46  &   7.35  &   7.47  &   7.89  &  19.32  &  19.85\\
   \cmidrule(lr){2-3}  \cmidrule(lr){4-6}  \cmidrule(lr){15-17} \cmidrule(lr){18-19}
   V$_{s}$\tablefootmark{c} (km s$^{-1}$)   &\multicolumn{2}{c}{-19.50}    & \multicolumn{3}{c}{-7.59}  &  &  &    &    &     &     &    &     &   \multicolumn{3}{c}{7.60}     & \multicolumn{2}{c}{19.50}  \\
\hline
\end{tabular}
\raggedright
\tablefoottext{a}{The hyperfine intensities are proportional to the square of the electric dipole moment matrix elements which are summed over the degenerate magnetic states of the transition, see \citet{1964JCP...40...257T}.\\}
\tablefoottext{b}{Relative to $\nu$ = 23694.4955 kHz, see \citet{1977ApJ...215L..35R}.\\}
\tablefoottext{c}{the hyperfine component intensity weighted velocity of the satellite.}
\end{sidewaystable*}

\subsection{Comparison with the observed spectra}
\label{intANDpeak}

Complementing Fig. \ref{fig:mspec}, we also present an example of an observed NH$_{3}$ (1,1) spectrum with a fitted velocity dispersion of 0.495 km s$^{-1}$ in Fig. \ref{fig:obspec} using the procedure outlined in Appendix \ref{hiaPro}.
The black polyline  shows the observed  NH$_{3}$ (1,1) spectrum.
The thick blue vertical lines below the zero level indicate the integrated ranges used to calculate the integrated HIA$_{\rm IS}$ and HIA$_{\rm OS}$.
The orange vertical lines indicate the ranges of the sub-spectra used to fit
the inner and outer satellite lines with single Gaussian functions.
The green line presents the result of the 18 hyperfine components fitting.
The red lines illustrate  the single Gaussian fitting results.
The cyan vertical lines show the 18 hyperfine components given in Table \ref{tab:nh3ratio}. The lengths and separations of these lines represent their predicted intensities and velocity separations.
Because the fitting methods are different and the satellite lines have intensity anomalies and asymmetric profiles, there are slight differences between the green (18 hyperfine components fitting) and red (single Gaussian fitting) lines.

In Fig. \ref{fig:compIP}, direct comparisons of the peak and integrated HIAs are presented.
The peak HIA$_{\rm IS}$ and integrated HIA$_{\rm IS}$ are shown in the left panel. The peak HIA$_{\rm OS}$ and integrated HIA$_{\rm OS}$ are presented in the right panel.
We can see in this figure that the peak HIAs are still roughly proportional  to the integrated HIAs, especially the HIA$_{\rm OS}$.
In Fig. \ref{fig:compIP-dis}, the peak (red dots) and integrated (blue dots) HIAs are plotted against the velocity dispersions of ISLs (left panel) and OSLs (right panel). We can see that from large to small velocity dispersions, the peak HIAs and the integrated HIAs are gradually separated (see also Fig. \ref{fig:simulation}). However, the differences between the peak and integrated HIAs due to computation methods are contaminated by real intensity anomalies and spectral noise.

In order to clearly demonstrate the similarities and differences between the HIAs derived from the observations and simulations (Appendix \ref{method}), we provide  the subtraction of the peak and integrated HIAs derived from the observed data (red and blue dots) against their velocity dispersions in Fig. \ref{fig:comp-m-m}.
We also present the subtraction of the peak and integrated HIAs of the simulated spectra against their velocity dispersions as red and blue lines in Fig. \ref{fig:comp-m-m}.
We can see that the derived subtraction of the HIAs from the observed data shows the same trend as the subtraction of the HIAs from the simulations (Fig. \ref{fig:simulation}), that is, the deviations of the peak and integrated HIAs are getting larger and larger with decreasing velocity dispersion. Nevertheless, the differences between the observed peak and integrated HIA$_{\rm OS}$ (red dots) are slightly smaller than the simulated ones.
From the simulations and observations, the peak and integrated HIAs show the same dichotomies.
Due to our simulations (see Fig. \ref{fig:simulation}), we know that the integrated HIAs should be more objective than the peak HIAs, especially dealing with spectra with low velocity dispersions, e.g spectra from quiescent clouds.

According to our results the commonly used definition of the HIA, i.e.  the peak HIA \citep[e.g.][]{2007MNRAS.379..535L, 2015ApJ...806...74C}, might be misleading.
Nevertheless, the peak HIAs are still roughly proportional to the integrated HIAs. Meanwhile, most of the sample  in \citet{2015ApJ...806...74C} are high-mass star formation regions, that means the spectra in that sample have considerable velocity dispersions (see their Fig. 3). In such objects   the difference between the peak and integrated HIAs is generally small. That is why the overall statistical results of our integrated HIAs are similar to the results of \citet{2015ApJ...806...74C}.

\begin{figure}
\includegraphics[width=\hsize]{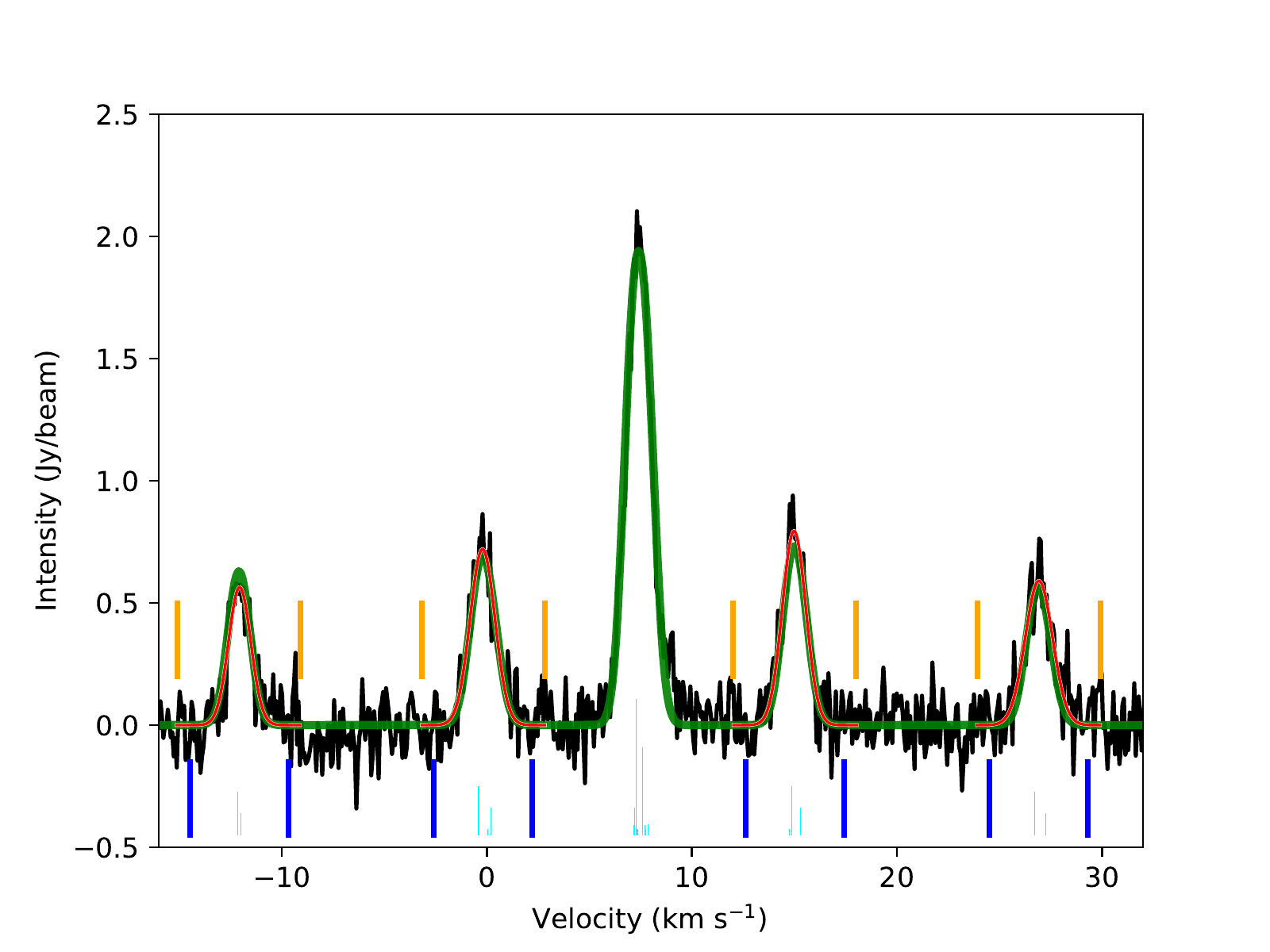}
\caption{An example of an observed  NH$_{3}$ (1,1) spectrum with a fitted velocity dispersion of 0.495 km s$^{-1}$. The black line presents the observed spectrum.  The green line presents the result of the 18 hyperfine components  fitting.
The thick orange vertical lines show the ranges of the sub-spectra used to fit the inner and outer satellite lines with single Gaussian functions.
The thick blue vertical lines below the zero level show the ranges used to calculate the integrated hyperfine intensity anomalies of the inner and outer satellite lines.
The red lines indicate the single Gaussian fitting results. The cyan vertical lines below the zero level present the 18 hyperfine individual components listed Table \ref{tab:nh3ratio}. The lengths and separations of these cyan lines  denote their expected intensities and velocity separations.}
\label{fig:obspec}
\end{figure}

\begin{figure*}
\centering
\includegraphics[width=0.45\textwidth]{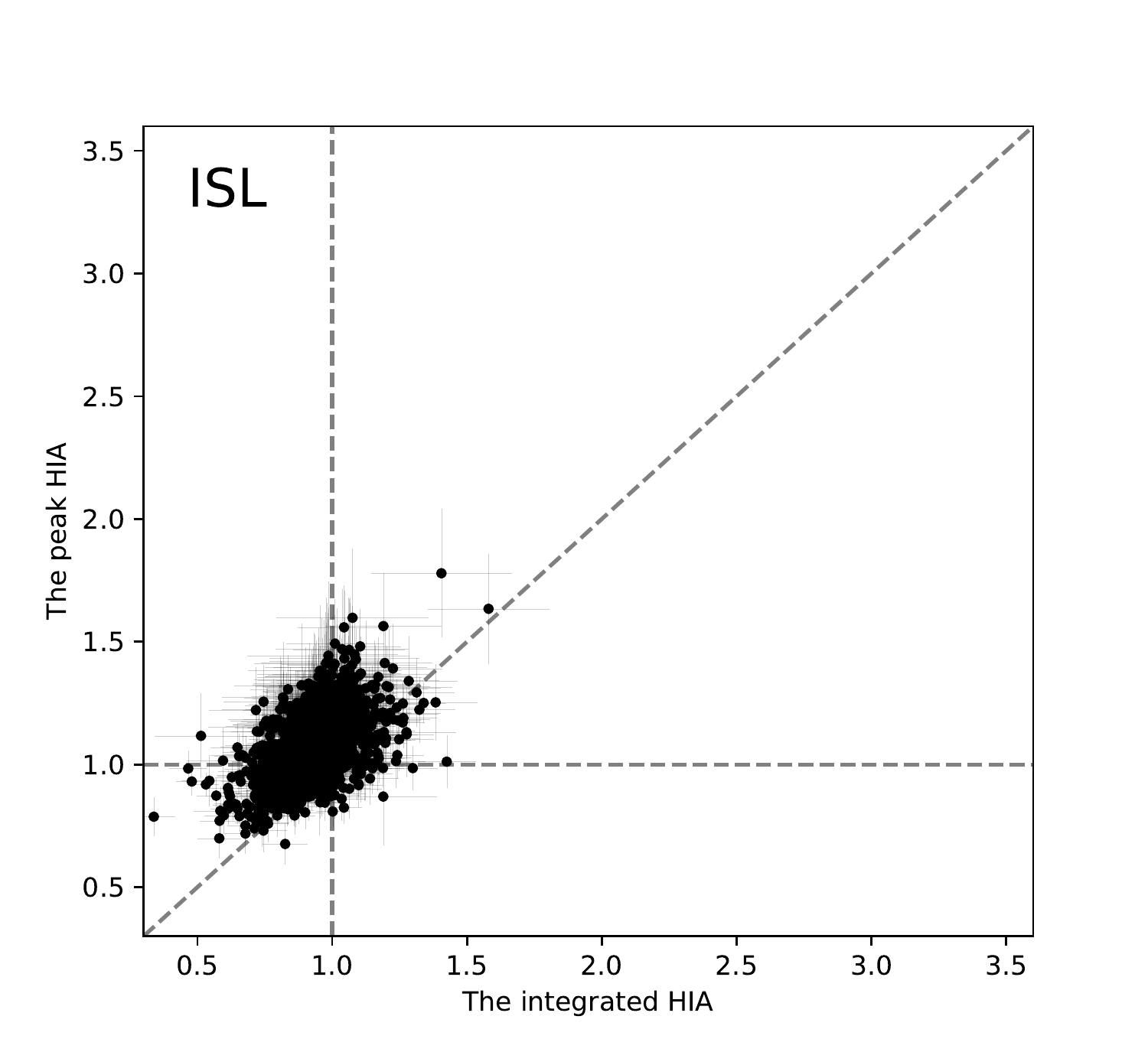}
\includegraphics[width=0.45\textwidth]{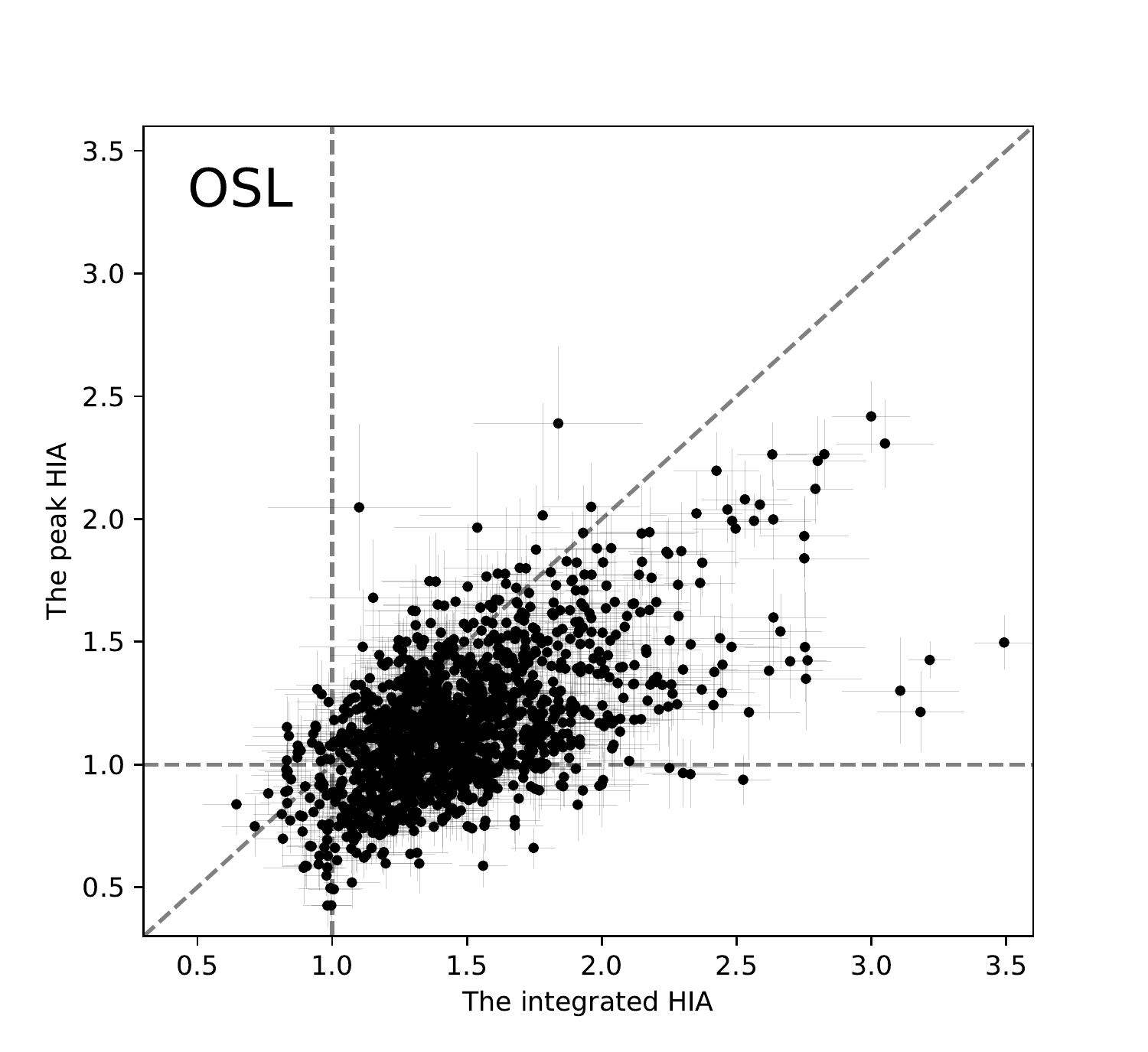}
\caption{The peak hyperfine intensity anomalies against the integrated hyperfine intensity anomalies of the inner satellite lines (left panel) and outer satellite lines (right panel).}
\label{fig:compIP}
\end{figure*}

\begin{figure*}
\centering
\includegraphics[width=0.45\textwidth]{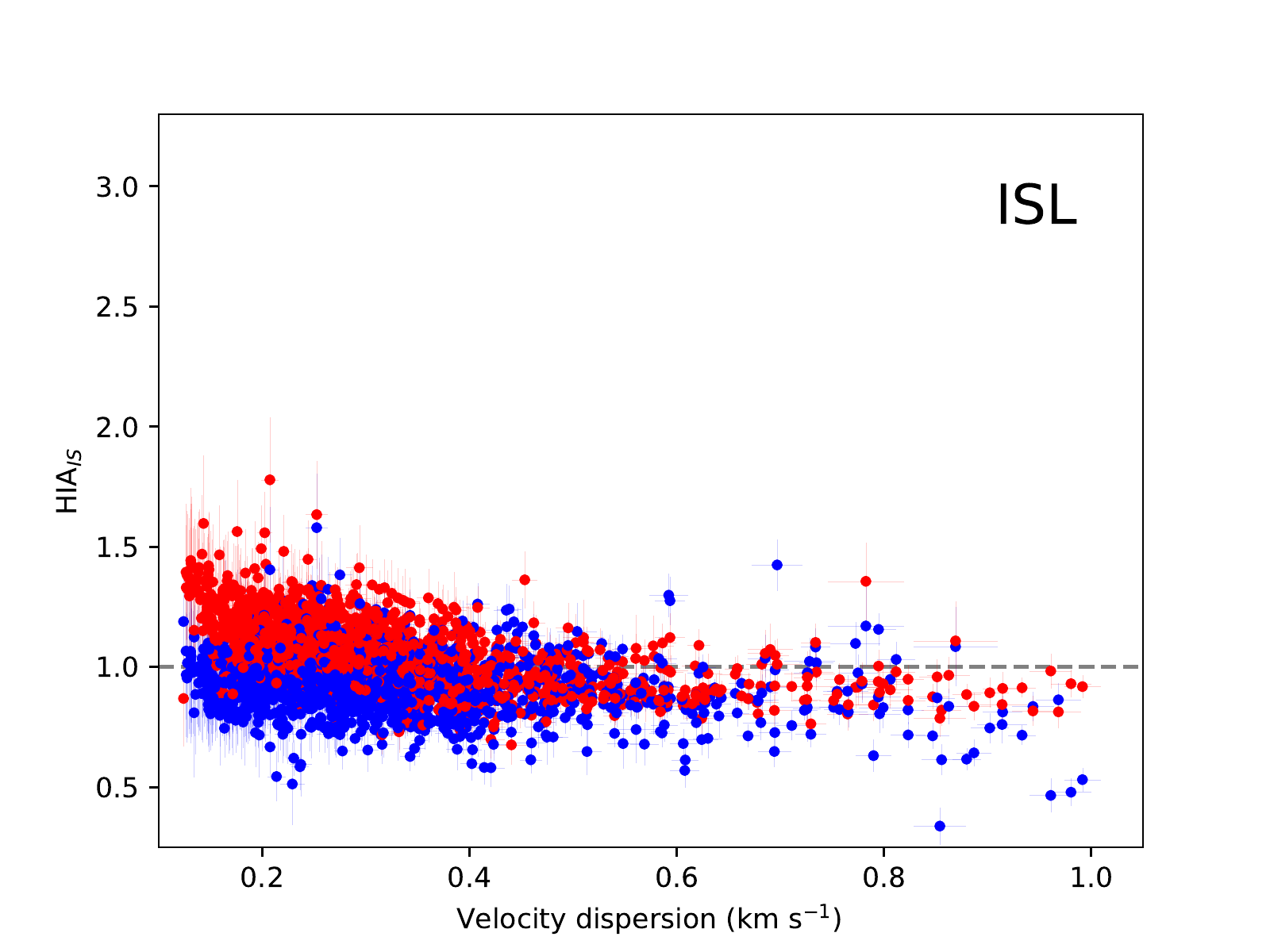}
\includegraphics[width=0.45\textwidth]{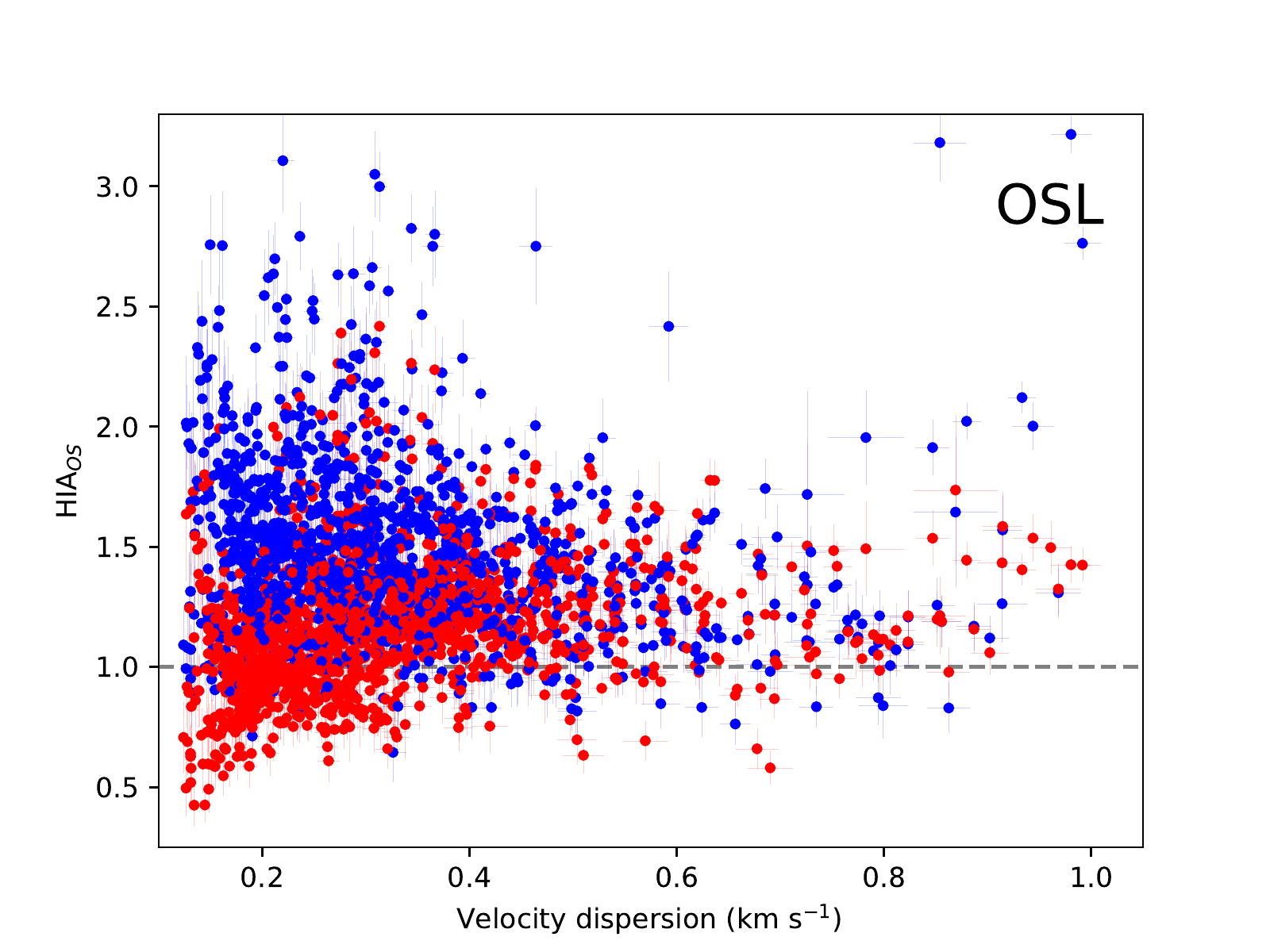}
\caption{The peak (red) and integrated (blue) hyperfine intensity anomalies of the inner satellite lines (left panel) and outer satellite lines (right panel) against the velocity dispersion.}
\label{fig:compIP-dis}
\end{figure*}

\begin{figure*}
\includegraphics[width=\hsize]{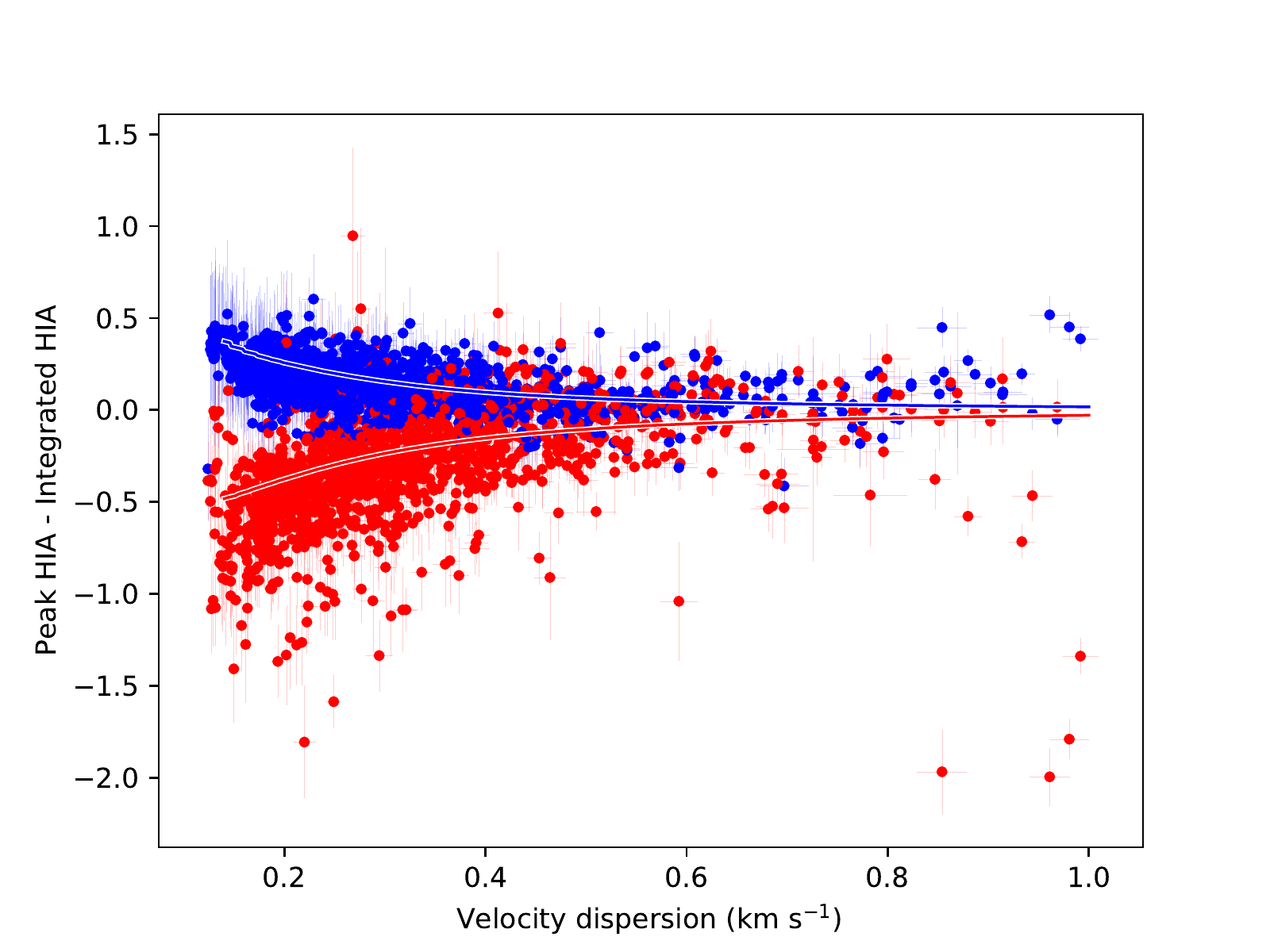}
\caption{The subtraction of the peak and the integrated hyperfine intensity anomalies of the inner satellite lines (blue dots) and outer satellite lines (red dots) against the velocity dispersion. The blue and red lines denote the subtraction of the peak and integrated hyperfine intensity anomalies as obtained from  the simulations presented in Fig. \ref{fig:simulation}.}
\label{fig:comp-m-m}
\end{figure*}

\section{The HIA procedure}
\label{hiaPro}

In order to quantitatively investigate the integrated HIAs within the Orion A MC, we developed an HIA procedure to  fit the NH$_{3}$ (1,1) spectra and calculate the integrated and also the peak HIAs by the following steps:

\begin{enumerate}
  \item Extract the spectra with a single velocity component and an signal-to-noise ratio (SNR) larger than 15. The methods are illustrated in \citet{2018RAA....18...77W}, that is, (1) the integrated intensity over the main line, i.e. the central group
        of hyperfine components, must be larger than 15 times the noise (the spectral RMS noise times the square root of the number of channels covering the main line); (2) within the main line, there must be a minimum of two adjacent spectral channels also showing emission larger than 15 times their spectral RMS noise.

  \item We used the combined 18 Gaussian functions with the fixed relative amplitudes and velocity separations listed in Table \ref{tab:nh3ratio} to fit the observed/simulated NH$_{3}$ (1,1) spectra.
      The observed data are fitted using optical depth as a function of velocity over the spectral band as in equation (A.2),

      \begin{equation}
      S(V) = A_{0} \times (1-e^{-\tau(V)}).
      \end{equation}

      The optical depth  of each  hyperfine component is assumed to have a Gaussian profile and to be proportional to their predicted intensities in case of LTE \citep[e.g.][]{2009ApJ...697.1457F, 2017PASP..129b5003E}.
      \begin{equation}
      \tau(V) = \tau_{0} \sum^{18}_{i=1} R_{\rm hfs}(i) \times e^{-(V - V_{0} - V_{\rm hfs}[i])^{2} /(2 \times \sigma_{\rm V}^{2})}.
      \end{equation}

      The free parameters are the amplitude and velocity offset of the combined spectrum $A_{0}$ and $V_{0}$, the intrinsic velocity dispersion \citep[][]{1998ApJ...504..207B} of each of the 18 hyperfine components $\sigma_{\rm V}$, and the total optical depth of NH$_{3}$(1,1), $\tau_{0}$.  $R_{\rm hfs}$ and $V_{\rm hfs}$ denote  the predicted relative intensities in the optically thin case and velocity offsets of the 18 hyperfine components (see Table \ref{tab:nh3ratio}).

  \item Based on  the fitted velocity and the intrinsic velocity dispersion  derived in step 2, we define the integrated velocity ranges as given in equation B.3 (thick blue vertical lines below the zero level in Figs. \ref{fig:mspec} and \ref{fig:obspec}).
       These are used to calculate the integrated HIAs. For the four sub-components we use equation B.4 (thick orange vertical lines in Figs. \ref{fig:mspec} and \ref{fig:obspec}) to fit the ISLs and OSLs with single Gaussian functions to obtain the peak HIAs. Here we applied two factors, $f_{\rm int}$ and  $f_{\rm peak}$ with
       \begin{equation}
        V_{0} + V_{s} \pm f_{\rm int} \times \Delta V
       \end{equation}
    and
        \begin{equation}
        V_{0} + V_{s} \pm f_{\rm peak} \times \Delta V.
      \end{equation}

       The hyperfine component intensity weighted velocity of the satellite is  $V_{s} = \sum^{i} R_{\rm hfs}[i] \times V_{\rm hfs}[i] / \sum^{i} R_{\rm hfs}[i]$ which is also given in Table \ref{tab:hia-params}.
       $\sum^{i}$ indicates that the summation is only for the two or three hyperfine components within each of the satellite lines. The FWHM line width $\Delta V = \sqrt{8 \times \ln(2)} \times \sigma_{\rm V}$.
       $V_{0}$ represents the fitted velocity offset of the combined spectrum, $\sigma_{\rm V}$ represents the fitted velocity dispersion of each of the 18 hyperfine components. $V_{0}$ and $\sigma_{\rm V}$ are all derived in step 2.
       The factor $f_{\rm peak}$ (equation A.5) is set to 2.5. The factor $f_{\rm int}$ (equation A.4) is set to 2.
       This is because in our simulations, this choice largely restrains fluctuations and oscillations (see Section \ref{method}) in the velocity dispersion range of 0.14--1.0 km s$^{-1}$ (the range of the considered velocity dispersions).
       The velocity range considered for the derivation of the integrated HIA has been chosen to be slightly narrower than the one for the peak HIA to        minimize the influence of noise in the calculation of integrated HIA values (see equation B.5).

  \item For the integrated HIAs, we calculate the integrated intensities and the integrated HIA$_{\rm IS}$ and HIA$_{\rm OS}$. For the peak HIAs, we fit each of the four satellite lines (sub-spectra) with a single Gaussian function and calculate the peak HIA$_{\rm IS}$ and HIA$_{\rm OS}$.
      We should note that this peak HIA calculation method is the same as that used in e.g. \citet[][]{1984A&A...139..258S}, \citet[][]{2007MNRAS.379..535L} and \citet[][]{2015ApJ...806...74C}

  \item The standard deviations of the integrated and peak HIAs  ($\sigma_{\rm int}$, $\sigma_{\rm peak}$) are assigned by

    \begin{equation}
        \sigma_{\rm int} = HIA_{\rm int} \times \sigma_{\rm hfs} \times \sqrt{ N_{\rm c} /(F_{\rm red} )^{2} +  N_{\rm c} /(F_{\rm blue} )^{2} },
    \end{equation}

    \begin{equation}
        \sigma_{\rm peak} = HIA_{\rm peak} \times \sqrt{( \sigma_{\rm peak, red}/P_{\rm red})^{2} + ( \sigma_{\rm peak, blue}/P_{\rm blue})^{2}}  ,
    \end{equation}

  where, $HIA\rm_{peak}$ and $HIA\rm_{int}$ are peak and integrated HIAs. $N\rm_{c}$ is the channel number within the integrated range. $\sigma\rm_{hfs}$ is  the standard deviation of the 18 hyperfine components fitting (step 2).
  $F\rm_{red}$ and $F\rm_{blue}$ are the integrated intensities of the redshifted  and blueshifted sides  of the  ISLs and OSLs.
  $\sigma_{\rm peak,red}$ and $\sigma_{\rm peak,blue}$ are the standard deviations of the single Gaussian fittings (step 4) of the redshifted and blueshifted sides of the ISLs or OSLs, respectively. $P\rm_{red}$ and $P\rm_{blue}$ are the fitted peaks of the blueshifted and redshifted sides  of the ISLs or OSLs, respectively.

\end{enumerate}

\end{appendix}

\end{document}